\documentclass[twocolumn,trackchange]{aastex62}

\graphicspath{{./}{figures/}}

\accepted{for publication in ApJS}

\shorttitle{The AstroSat UV Deep Field north}
\shortauthors{Mondal et al.}

\begin{document}

\title{The AstroSat UV Deep Field North: the far and near ultraviolet photometric catalog}

\author[0000-0003-4531-0945]{Chayan Mondal}
\affiliation{Inter-University Centre for Astronomy and Astrophysics, Ganeshkhind, Post Bag 4, Pune 411007, India}

\email{chayanm@iucaa.in, mondalchayan1991@gmail.com}

\author[0000-0002-8768-9298]{Kanak Saha}
\affiliation{Inter-University Centre for Astronomy and Astrophysics, Ganeshkhind, Post Bag 4, Pune 411007, India}

\author[0000-0003-4594-6943]{Souradeep Bhattacharya}
\affiliation{Inter-University Centre for Astronomy and Astrophysics, Ganeshkhind, Post Bag 4, Pune 411007, India}

\author[0000-0002-2870-7716]{Anshuman Borgohain}
\affiliation{Department of Physics, Tezpur University, Napaam, India}

\author[0000-0001-6350-7421]{Shyam N. Tandon}
\affiliation{Inter-University Centre for Astronomy and Astrophysics, Ganeshkhind, Post Bag 4, Pune 411007, India}

\author[0000-0002-9946-4731]{Marc Rafelski}
\affiliation{Space Telescope Science Institute, 3700 San Martin Drive, Baltimore, MD 21218, USA}
\affiliation{Department of Physics and Astronomy, Johns Hopkins University, Baltimore, MD 21218, USA}
\author[0000-0003-1268-5230]{Rolf A.~Jansen}
\affiliation{School of Earth \& Space Exploration, Arizona State University, Tempe, AZ 85287-1404, USA}
\author[0000-0001-8156-6281]{Rogier A. Windhorst}
\affiliation{School of Earth \& Space Exploration, Arizona State University, Tempe, AZ 85287-1404, USA}
\author[0000-0002-7064-5424]{Harry I. Teplitz}
\affiliation{IPAC, Mail Code 314-6, California Institute of Technology, 1200 E. California Blvd., Pasadena CA, 91125, USA}
\author[0000-0002-0648-1699]{Brent M. Smith}
\affiliation{School of Earth \& Space Exploration, Arizona State University, Tempe, AZ 85287-1404, USA}

\begin{abstract}
We present deep UV imaging observations of the Great Observatories
Origins Survey Northern (GOODS-N) field with AstroSat/UVIT (AstroSat UV Deep Field north - AUDFn), using one far-UV (F154W, 34.0\,kilosec) and two near-UV filters (N242W, 19.2\,kilosec; N245M, 15.5\,kilosec). The nature of the UV sky background was explored across the UVIT field and a global mean and rms was estimated for each filter. We reach 3$\sigma$ detection limits of $m_{\rm AB}$ $\sim$ 27.35~mag, 27.28~mag and 27.02~mag for a point source in the F154W, N242W and N245M bands respectively. The 50\% completeness limits of the FUV and NUV images are $m_{\rm AB}=$ 26.40~mag and 27.05~mag respectively. We constructed PSFs for each band and estimated their FWHM, which were found to be almost the same: 1\farcs18 in F154W, 1\farcs11 in N242W, and 1\farcs24 in N245M. We used SExtractor to separately identify sources in the FUV and NUV filters and produce the UV source catalog of the entire AUDFn field. The source count slope estimated in FUV and NUV is 0.57 dex mag$^{-1}$ (between 19 - 25 mag) and 0.44 dex mag$^{-1}$ (between 18 - 25 mag), respectively. The catalog contains 6839 and 16171 sources (brighter than the 50\% completeness limit) in the FUV and NUV, respectively. Our FUV and NUV flux measurements of the identified sources complement existing multi-band data in the GOODS-N field, and enable us to probe rest-frame FUV properties of galaxies at redshift $z < 1$ and search for candidate Lyman continuum leakers at redshift $z > 0.97$.

\end{abstract}

\section{Introduction}
The multi-band photometric observation of deep fields has offered the unique scope to identify and characterize large number of galaxies at varying redshifts starting from cosmic dawn to recent times. The wide wavelength coverage from ultra-violet (UV) to infra-red (IR) has been utilised to build SEDs of galaxies to estimate their stellar mass, star formation rate (SFR), star formation history (SFH) and photometric redshift. These rich galaxy samples are used to study the redshift evolution of the stellar mass function (e.g. \citet{perez2008,marchesini2009,ryan2007}) as well as their structural and morphological properties \citep{franx2008,bell2012,vanderwel2012,bouwens2004,oesch2010}. The Wide Field Camera 3 (WFC3) and Advanced Camera for Surveys (ACS) instruments of the Hubble Space Telescope (HST) have played a major role in producing deep, high-resolution images of several deep fields with different targeted surveys, such as the Hubble Ultra Deep Field (HUDF; \citet{beckwith2006}), and the Cosmic Assembly Near-infrared Deep Extragalactic Legacy Survey (CANDELS; \citet{grogin2011,koekemoer2011}) which observed five fields (AEGIS, COSMOS, GOODS-North, GOODS-South, and UDS) with multiband imaging. Each field in CANDELS has further been observed by several ground (e.g., CFHT, KPNO, Subaru) as well as space (e.g., Chandra, Spitzer) telescopes \citep{skelton2014}. Despite having such a rich data sample, these fields lack high-resolution deep far-UV (FUV) and near-UV (NUV) observations for the entire area as covered with optical and infra-red bands. The Galaxy Evolution Explorer (GALEX; \citet{martin2005}) telescope has covered these fields in the FUV and NUV, but due to its poor angular resolution (FWHM\,$\sim$\,5\farcs3) it remains less effective for deep field studies. The UVIS channel of the HST WFC3 camera has provided superior NUV images of several deep fields (e.g., the Hubble Ultraviolet Ultra Deep Field (UVUDF; \citealt{teplitz2013}), the Hubble Deep UV Legacy Survey (HDUV; \citealt{oesch2018}) and the recent UVCANDELS survey (PI: H.~Teplitz), but observations to similar depth in the FUV band covering the entire deep field remain lacking. 
Hence, with this motivation, we conduct targeted UV imaging of the AstroSat UV Deep Field south (AUDFs) and north (AUDFn) that fully encompass two CANDELS fields, GOODS-south and GOODS-north, using both FUV and NUV channels of the Ultra-Violet Imaging Telescope
\citep[UVIT; ][]{kumar2012} and have FWHM $\sim$ 1\farcs4{\,}.

In this paper, we present the UVIT FUV and NUV observations of the GOODS-N field \citep{giavalisco2004}. GOODS-N is a large patch in the sky encompassing the Hubble Deep Field North (HDFN \citealt{williams1996}) in its central region. The field has been extensively observed with the HST WFC3/UVIS (F275W, F336W), ACS/WFC (F435W, F606W, F775W, F850W, F814W, F850LP) and WFC3/IR (F275W, F336W, F125W, F140W, F160W) cameras to cover $\sim$ 164 square arcmin area centred on
(RA, Dec)$_{\rm J2000}$ = (12:36:55, +62:14:15)
\citep{giavalisco2004,grogin2011,koekemoer2011,skelton2014,oesch2018}. Targeted observations using HST FUV channels are also conducted for specific part of the GOODS-N field \citep{teplitz2006}. The field has also been observed with other telescopes
over a wide wavelength range from X-ray through radio: extant data include deep X-ray observations by \textit{Chandra} \citep{alexander2003}, \textit{U} band imaging using the Kitt Peak 4-m telescope \citep{capak2004}, optical \textit{BVRiz} observations from the Subaru 8.2-m \citep{capak2004} and \textit{G, R$_s$} from Keck\,I \citep{steidel2003}, 25 optical medium bands from the 10.4-m GranTeCan \citep{perez2013}, near-infrared \textit{J, H, K$_s$} from Subaru \citep{kajisawa2011}, 3.6 and 4.5\,$\mu$m \citep{ashby2013} and 5.8, 8, 24, and 70\,$\mu$m \citep{dickinson2003} from \textit{Spitzer}, 100--500\,$\mu$m far-infrared observations from \textit{Herschel} \citep{elbaz2011,magnelli2013}, and 1.4 and 5.5\,GHz radio observations from the VLA \citep{morrison2010,guidetti2017}.
Apart from these, several spectroscopic surveys have also been conducted in GOODS-N, which provide redshifts for many identified galaxies. The most extensive compilations of spectroscopic measurements include the AGHAST survey using HST's WFC3/IR G141 grism (PI - B. Weiner) and  ACS G800L grism (PI - G. Barro) (both incorporated in the 3D-HST survey by \citet{momcheva2016}), and Keck observations from the TKRS2 survey using MOSFIRE \citep{wirth2015}, LRIS and DEIMOS spectra \citep{reddy2006,barger2008}, the DEEP3 galaxy redshift survey using DEIMOS \citep{cooper2011}, and rest-frame optical spectra from the MOSDEF survey \citep{kriek2015}.
This photometric and spectroscopic database has resulted in several important studies on galaxies covering a wide redshift range (e.g., \citet{kobulnicky2004,reddy2004,grazian2017}).
Nonetheless, due to the unavailability of high-resolution UV images (as the bluest HST filter used for GOODS-N is F275W), galaxies at $z$\,$<$\,1 still remain less explored. The UVIT observations presented in this study will fill this gap by providing the rest-frame FUV flux of galaxies between $z$\,$\sim$\,0 and $z$\,$\sim$\,1. The UVIT data will enable the study of the UV continuum slope $\beta$ (Mondal et al. in prep.), UV luminosity function, and SFR of galaxies within this redshift range. Furthermore, these observations will enable the search for potential LyC leaking galaxies in the intermediate redshift range, in which the first discovery has already been reported by \citet{saha2020} at redshift $z$\,=\,1.42 in GOODS-south.

Here, we present the details of UVIT observations including the preparation of deep science ready images from the raw data. We describe the image products in different filters, explore the nature of FUV and NUV background across the field and also construct PSFs. We use SExtractor \citep{bertin1996} to identify objects in FUV and NUV images and perform aperture photometry to estimate their flux. We provide catalogs for both FUV and NUV bands listing the position, size, and magnitude of detected sources located within 13.0$^{\prime}$ from the UVIT field centre. To understand the quality of our detection as well as measurement, we estimate the 50\% completeness limit and also the difference between input and retrieved magnitudes of artificially injected sources. Using the positional prior from the HST CANDELS/GOODS-N catalog of \citet{barro2019}, we produced another catalog listing the FUV and NUV magnitudes of the HST-detected clean sources. Our measurements add a valuable asset to the existing rich database of GOODS-N and also offer one of the first deep field observations conducted using UVIT. The paper is arranged as follows: we discuss the UVIT observations and data reduction, background estimation, and PSF modelling in \S\ref{s_data}, the source detection, photometry, catalog content, and completeness in \S\ref{s_source_detection}, followed by a summary in \S\ref{s_summary}. Throughout the paper, we provide magnitudes in the AB system \citep{okegunn83}.

\begin{figure*}
    \centering
    \includegraphics[width=6.5in]{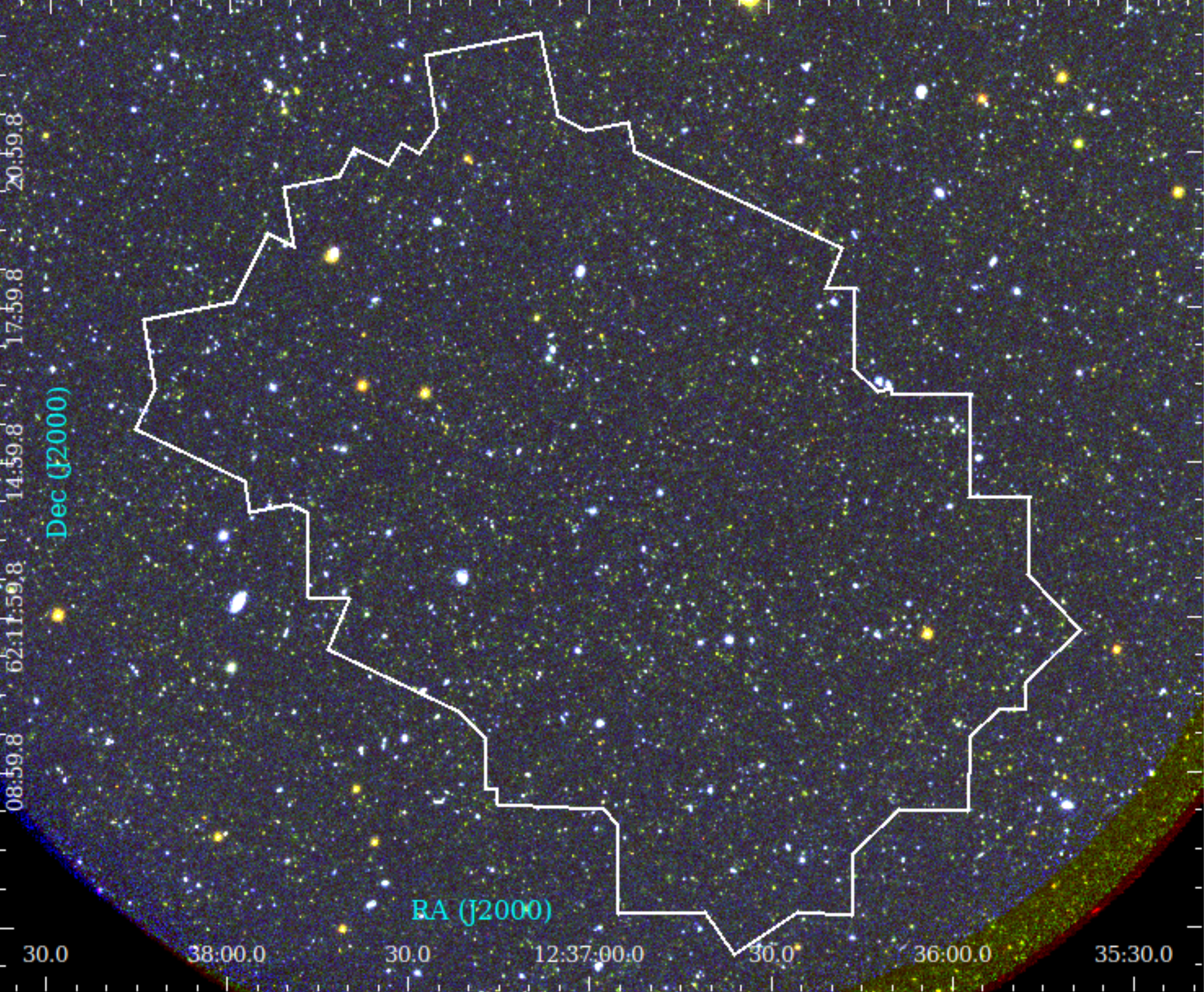} 
    \caption{False color composite image of the GOODS-N field observed by AstroSat/UVIT. The background shows a part of the full UVIT circular field of diameter 28 arcmin. The blue, red, and green colors represent emission in F154W, N242W, and N245M filters respectively. The white polygon shows the HST coverage as observed in CANDELS \citep{grogin2011,koekemoer2011}. The catalog presented in this work includes sources detected within 13.0$^{\prime}$ from the UVIT field centre (as shown in Figure \ref{fig_f154w_bg_map}).}
    \label{fig_color}
\end{figure*}

\section{Data and Observations}
\label{s_data}

\begin{figure}
    \centering
    \includegraphics[width=3.5in]{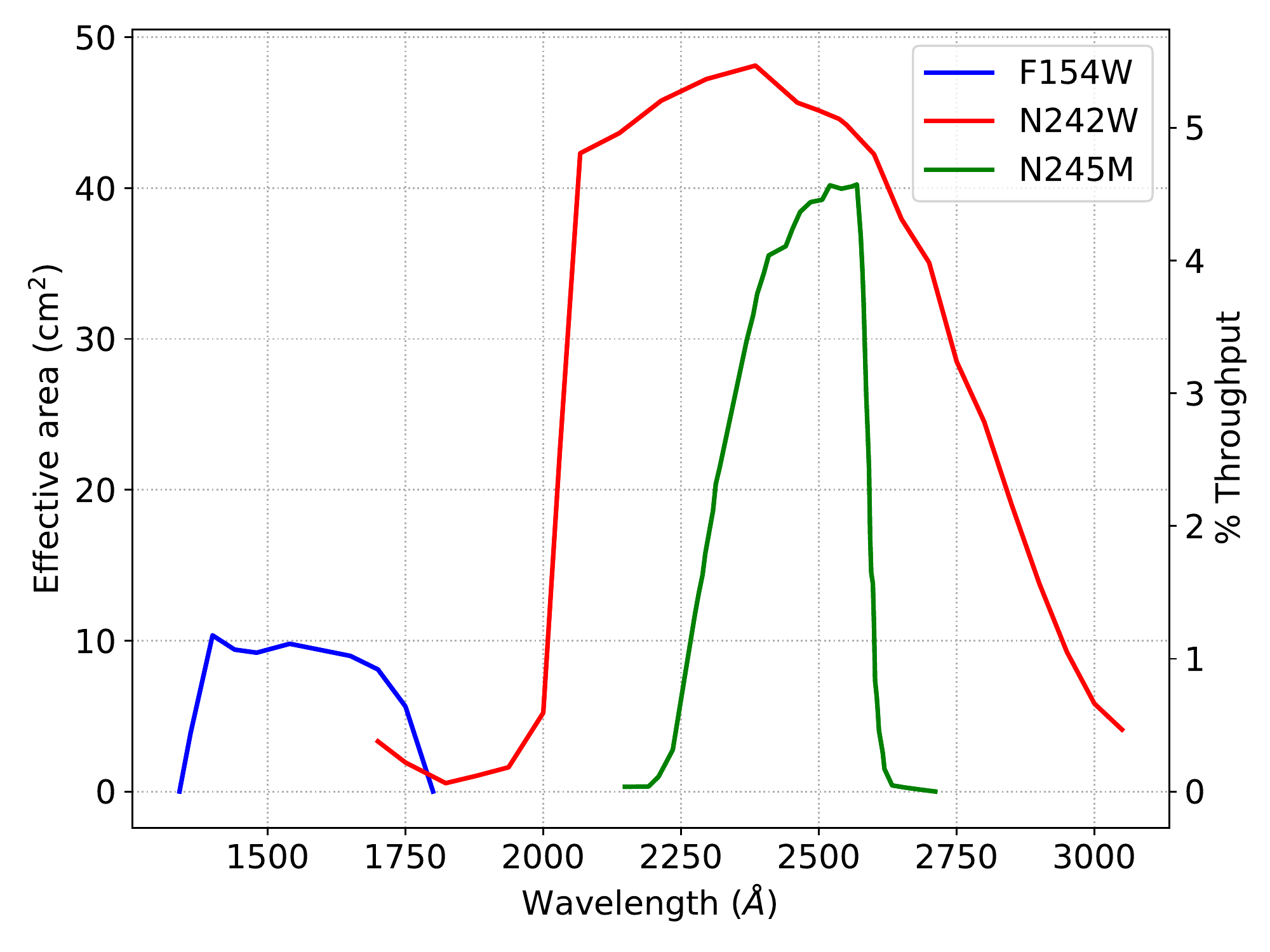} 
    \caption{Effective area profile of the three UVIT filters used for observation (data obtained from \citet{tandon2020}). The y-axis also shows the value of \% Throughput.}
    \label{fig_filters}
\end{figure}

\begin{figure}
    \centering
    \includegraphics[width=3.5in]{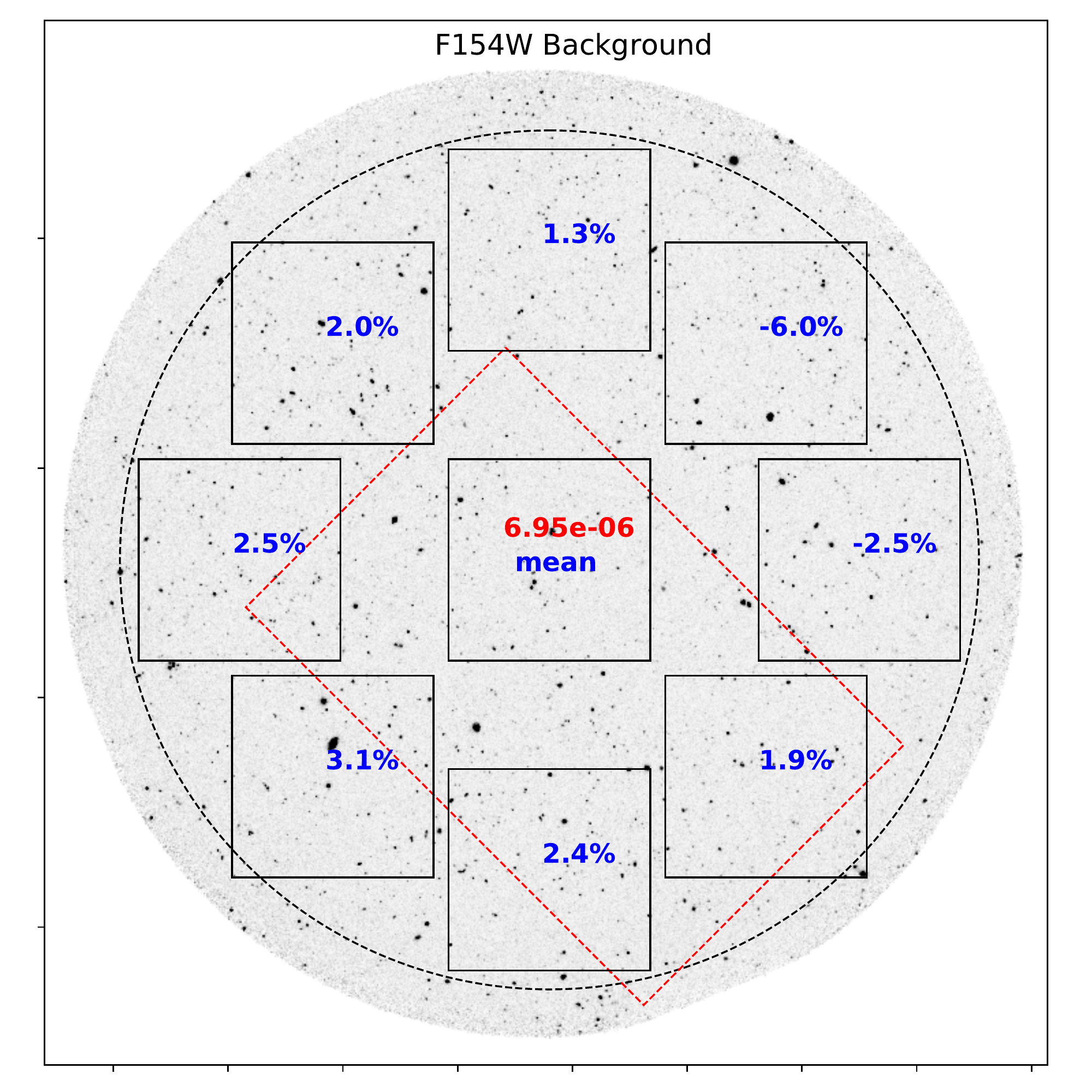} 
    \caption{The 6$^{\prime}\times$6$^{\prime}$ black squares show the 9 large boxes adopted to sample the background. The mean background value in cps/pixel estimated for the central box is shown in red. The numbers corresponding to other boxes (in blue) convey the measured percentage difference of the local background mean (i.e., measured within each box) from the estimated mean of the central box in one random run as explained in Section \ref{s_background_estimation}. The error value associated with the background flux within a 6$^{\prime}\times$6$^{\prime}$ box is $\sim$0.24\%.  The region outlined by the red dashed rectangle indicates the HST coverage. The black dashed circle, that has a radius of 13$^{\prime}$ from the UVIT field centre, shows the area for which we provide the catalog. This figure highlights the relative variation in the FUV background across the UVIT field.}
    \label{fig_f154w_bg_map}
\end{figure}

We used UVIT \citep{kumar2012} onboard the AstroSat \citep{singh2014} satellite  to observe the GOODS-N deep field. The UVIT instrument consists of two telescopes, each with a circular field of view with a diameter of 28$^{\prime}$. One telescope observes in a Far-ultraviolet (FUV) band while the other operates both a Near-ultraviolet (NUV) and a Visible channel. The instrument is capable of observing all three channels simultaneously, where FUV and NUV are used for scientific observations and the Visible channel is utilised mainly for tracking the drift pattern of the satellite pointing during the course of an observation. The detectors in the FUV and NUV channels work in photon counting mode (with a full-frame read rate of 28.7 Hz) where the centroid of each photon event provides the positional information \citep{postma2011}. Each of the three observing channels is equipped with multiple photometric filters of different bandwidths. We used two broad band filters (F154W for FUV and N242W for NUV) and one medium band filter (N245M for NUV) to observe the GOODS-N field. A UVIT color composite image of the observed field is shown in Figure \ref{fig_color}. Relevant details of the UVIT filters are listed in Table \ref{table_uvit_obs}. We show the effective area profile as a function of wavelength of these three filters in Figure \ref{fig_filters}. All the values related to filter calibration, including data for these effective area profiles can be found in the UVIT additional calibration paper by \citet{tandon2020}.
The full sequence of observations (proposal ID G08\_077; PI: Kanak Saha) was executed over 32 AstroSat orbits spanning UT 2018-03-10 to 2018-03-12. The HST CANDELS imaging \citep{barro2019,grogin2011,koekemoer2011} of the GOODS-N field covers a roughly rectangular area approximately 16$^{\prime}\times$10$^{\prime}$ in size, which UVIT could cover with a single pointing (Figure \ref{fig_color} \& Figure \ref{fig_f154w_bg_map}). Apart from the UVIT observations, we also make use of the HST CANDELS photometric catalog of the GOODS-N field from \citet{barro2019} in this work.\\

\begin{table*}
\caption{Details of the UVIT observations. The values related to filter calibration are obtained from \citet{tandon2020}.}
\label{table_uvit_obs}
\begin{tabular}{p{1.0cm}p{1.5cm}p{1.2cm}p{2.0cm}p{3.0cm}p{1.5cm}p{2.2cm}p{2.7cm}}
\hline
Filter & Bandpass & $\lambda_{mean}$ & ZP magnitude & Unit conversion & Exposure & 3$\sigma$ detection & 50\% completeness\\
 & (\AA) & (\AA) & (AB) & (erg sec$^{-1}$cm$^{-2}$\AA$^{-1}$) & time (sec) & limit (AB) m$_{3\sigma}$ & limit (AB)\\
 (1) & (2) & (3) & (4) & (5) & (6) & (7) & (8)\\\hline
F154W & 1340-1800 & 1541 & 17.771 & 3.57$\times10^{-15}$ & 34022.8 & 27.35 & 26.40\\
N242W & 1700-3050 & 2418 & 19.763 & 2.32$\times10^{-16}$ & 19228.6 & 27.28 & 27.05\\
N245M & 2148-2711 & 2447 & 18.452 & 7.57$\times10^{-16}$ & 15522.8 & 27.02 & 27.05\\\hline
\end{tabular}
\textbf{Note.} Table columns: (1) name of the UVIT filter used for observation; (2) filter bandpass in \AA; (3) filter mean wavelength in \AA; (4) filter zero point magnitude in the AB system; (5) filter unit conversion factor that signifies the energy (in CGS) of a single photon detected in each filter; (6) total exposure time of science images used for analysis; (7) 3$\sigma$ detection limit estimated using the background rms measured by SExtractor within a circular aperture of radius 1.0$^{\prime\prime}$ (equation \ref{e_detection}); (8) 50\% source completeness magnitude limit as estimated in Section \ref{sect:completeness} using artificially injected sources.
\end{table*}

\begin{figure}
    \centering
    \includegraphics[width=3.5in]{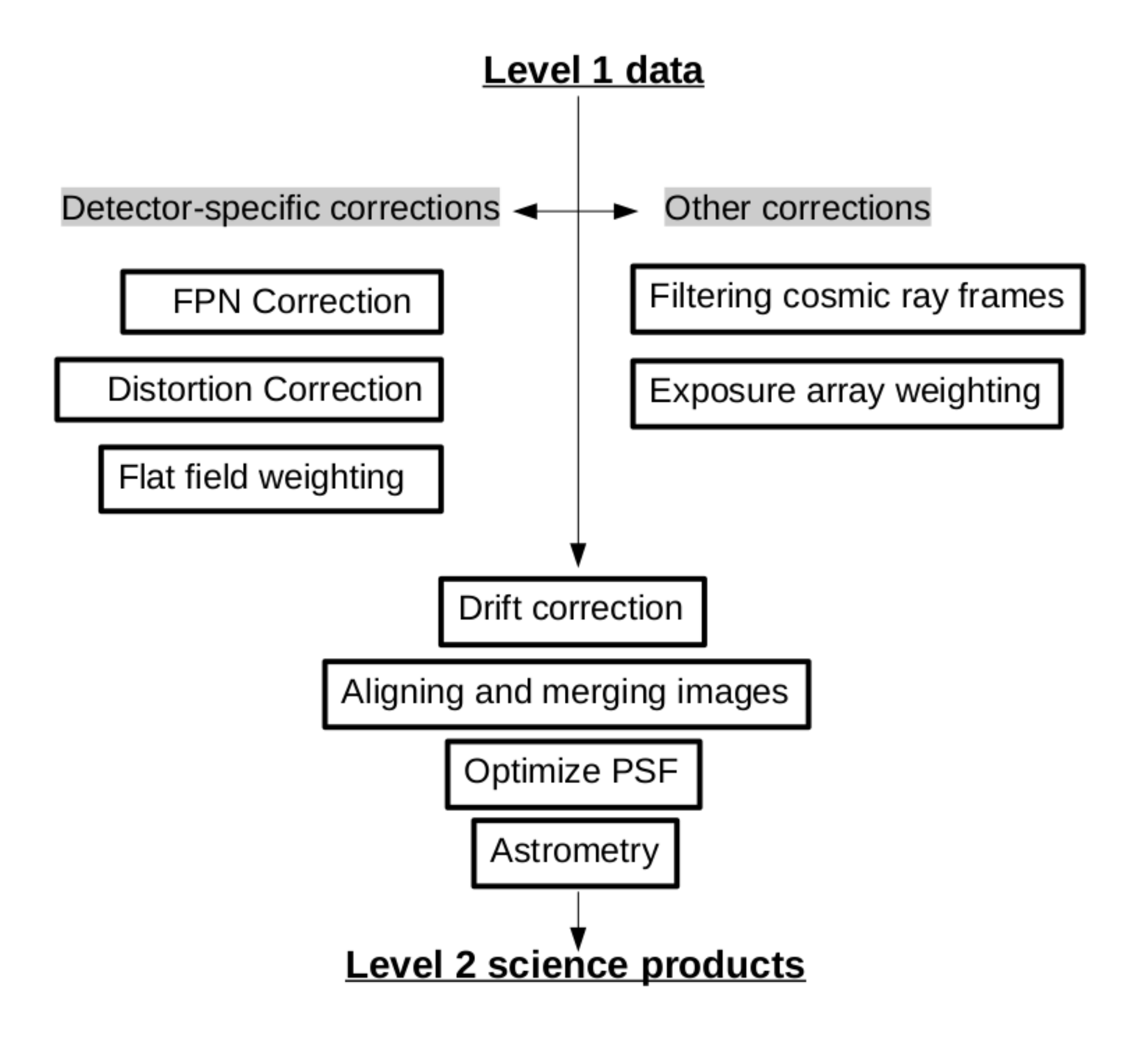} 
    \caption{Flowchart showing the data processing steps for producing UVIT Level 2 images from the raw Level 1 data using the CCDLAB software \citep{postma2017}. The calibration data for FPN noise, distortion, and flat fielding are obtained from the UVIT calibration database. Details of the data reduction procedure using CCDLAB are given in \citet{postma2021}.}
    \label{fig_ccdlab}
\end{figure}

\subsection{UVIT Level 1 data reduction}

We retrieved the raw Level~1 data of each orbit from the \textit{AstroSat Archive}\footnote{https://astrobrowse.issdc.gov.in/astro\_archive/archive/Home.jsp}
and used the public UVIT data reduction pipeline CCDLAB v3.0 \citep{postma2017,postma2021}
to produce science-ready images in each observed filter. The data reduction procedure
is illustrated in the flowchart of Figure~\ref{fig_ccdlab}. We apply detector-specific
corrections to counter the fixed pattern noise (FPN) and geometric distortions using the
UVIT calibration database from \citet{girish2017,postma2011}. The FPN calibration also
corrects for the systematic bias in centroid positions, and for distortions due to intrinsic
imperfections in the detector system.  Before merging individual data frames to produce
the Level~2 calibrated images, we removed frames that suffered cosmic ray hits.  After
a careful assessment of the appropriate threshold, we adopt a 4$\sigma$ cutoff to remove
frames that likely suffered from cosmic ray induced signal, where $\sigma$ = $\sqrt{I}$
and $I$ is the median number of counts/frame calculated over all observed frames.
Each remaining frame is flat fielded to correct for the spatially non-uniform sensitivity
of the detector. The resulting final exposure time for each band is listed in
Table~\ref{table_uvit_obs}.

Next, we apply a drift correction using the drift series information along the X and Y
detector axes from the Visible VIS3 observations, which were executed in parallel with
the observations in each of the three UV filters.  The drift corrected images of each
orbit were then aligned and co-added to produce deep mosaics in each filter. Using CCDLAB,
we further optimized the PSF in these deep images to correct for any higher-order drift
residuals \citep{postma2021}.  Each final image measures 4800$\times$4800 pixels
(including padding of areas without coverage) at a plate scale of $\sim$0\farcs417, and
was placed onto the Gaia EDR3 astrometric reference system \citep{gaia2021} using the
built-in algorithm in CCDLAB and sources detected by both Gaia and in these UV images.
The resulting astrometric accuracy of these science-ready images is $\sim$0\farcs25 (rms)
across the entire field.\\

\begin{figure*}
    \centering
    \includegraphics[width=2.3in]{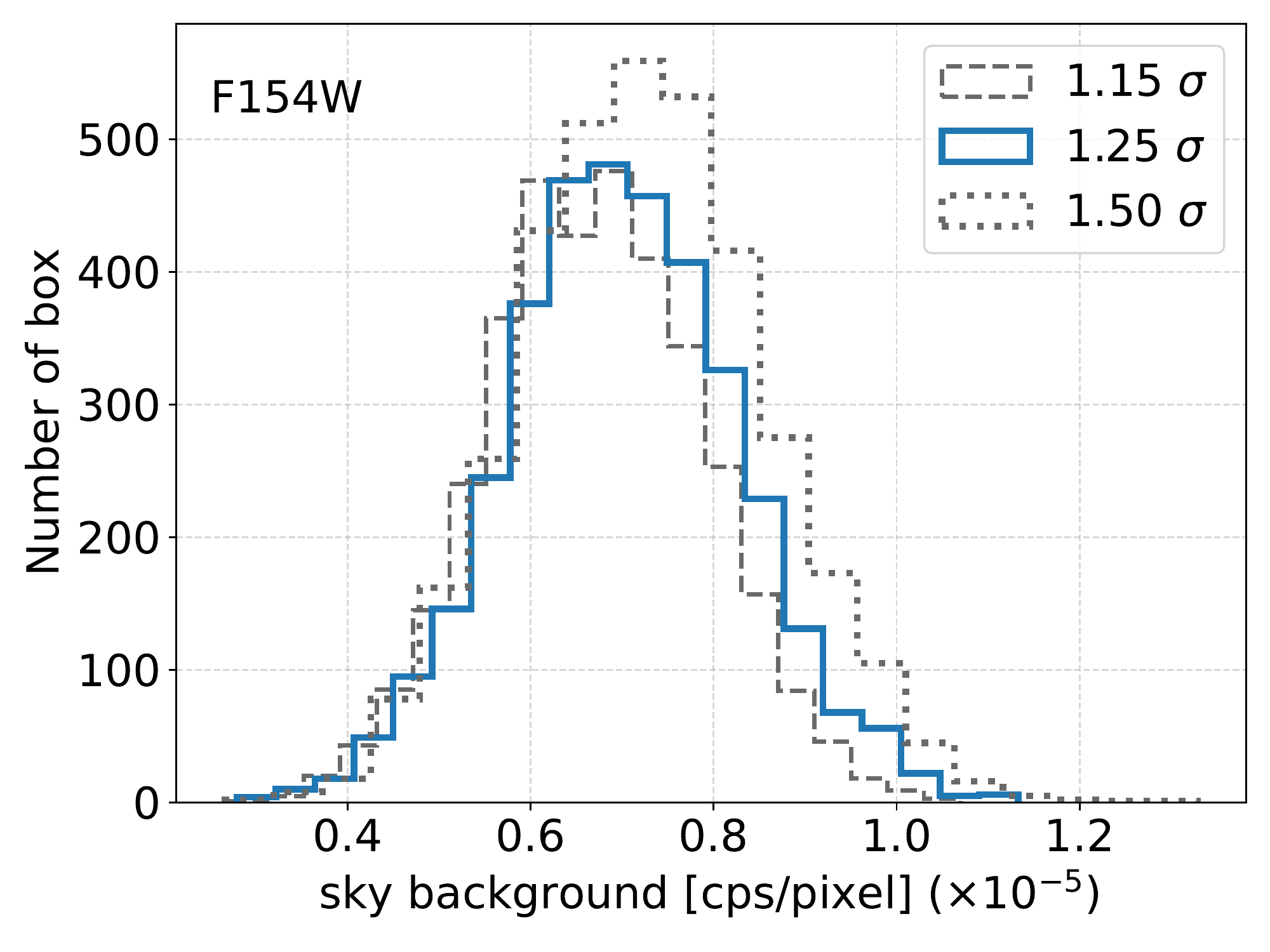}
    \includegraphics[width=2.3in]{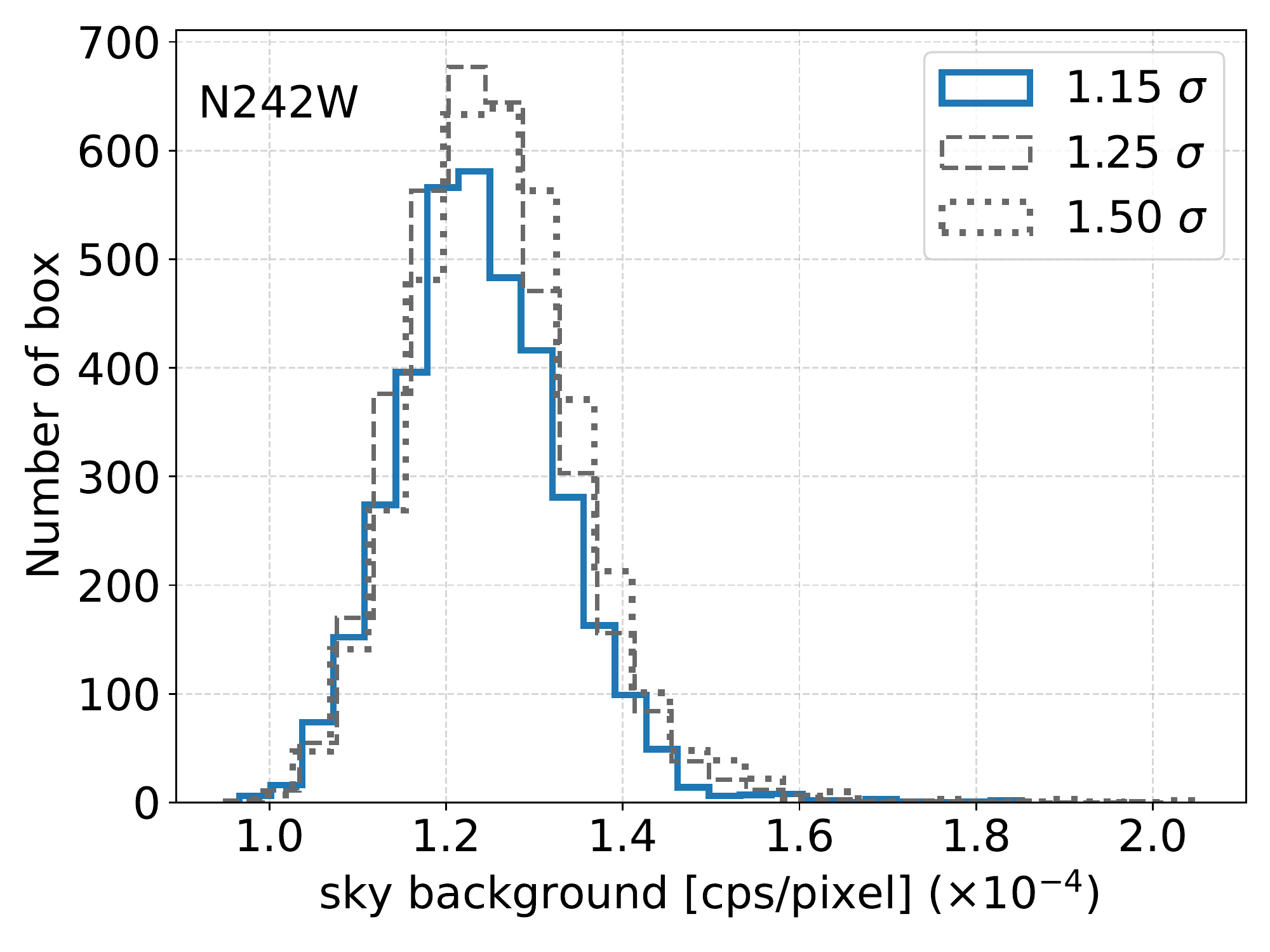}
    \includegraphics[width=2.3in]{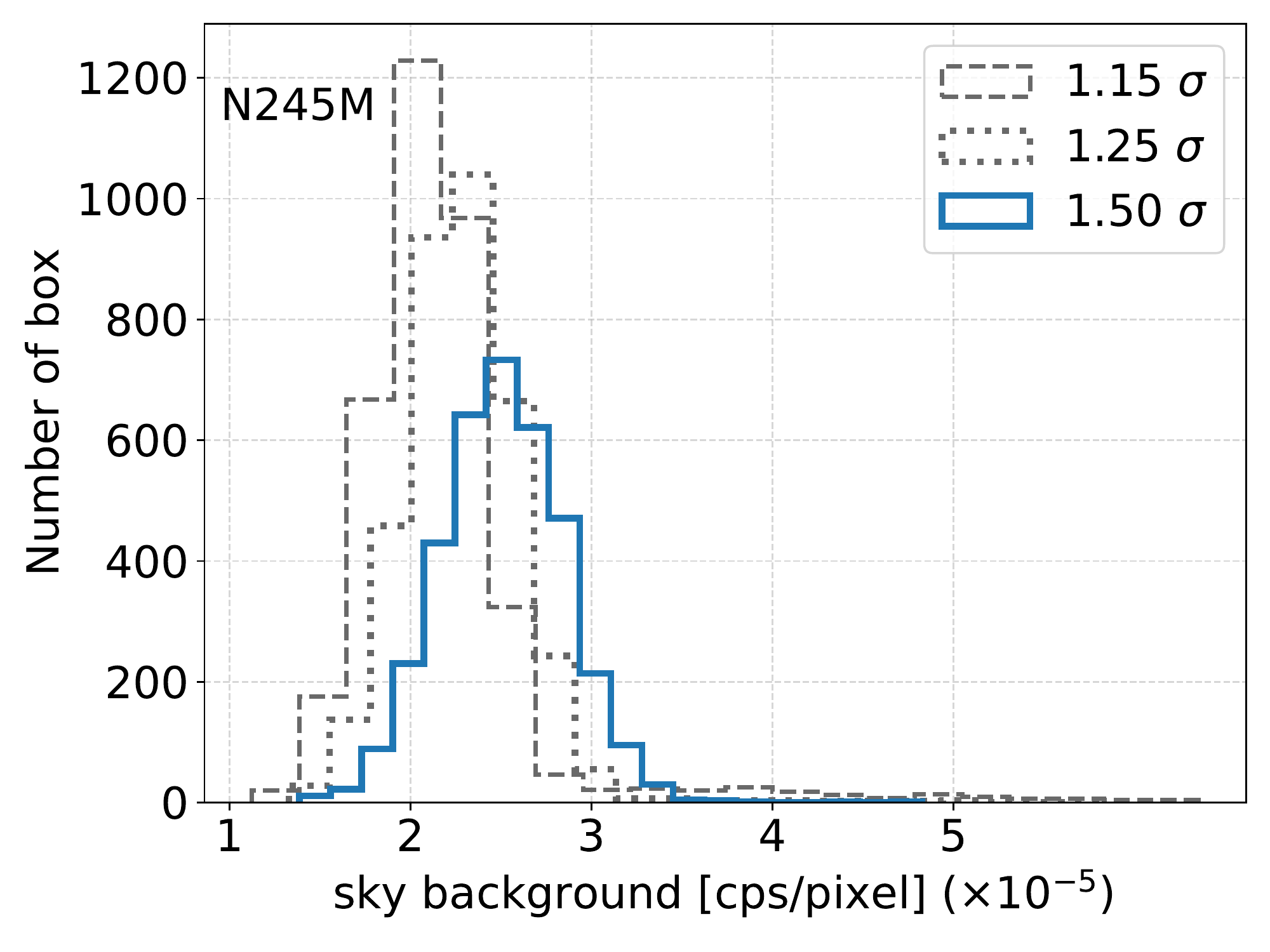}
    \caption{The distribution of the sky background as measured with 3600 random boxes (400 within each large box shown in Figure \ref{fig_f154w_bg_map}), each having a size of 11$\times$11 pixels in the UVIT images in F154W (\textit{left}), N242W (\textit{middle}) and N245M (\textit{right}). The optimal SExtractor detection thresholds adopted for F154W, N242W, and N245M filters are 1.25, 1.15, and 1.50$\sigma$ respectively. The histogram for the optimal threshold in each respective filter is shown in blue solid line, while the distributions for the other two thresholds are displayed (to show the variation in skewness) in grey dashed and dotted lines. For the optimal detection threshold, the plotted distribution shows one among the 100 random iterations performed to estimate the global background values listed in Table \ref{table_bg}.}
    \label{fig_f154w_bghist}
\end{figure*}

\begin{figure*}
    \centering
    \includegraphics[width=2.3in]{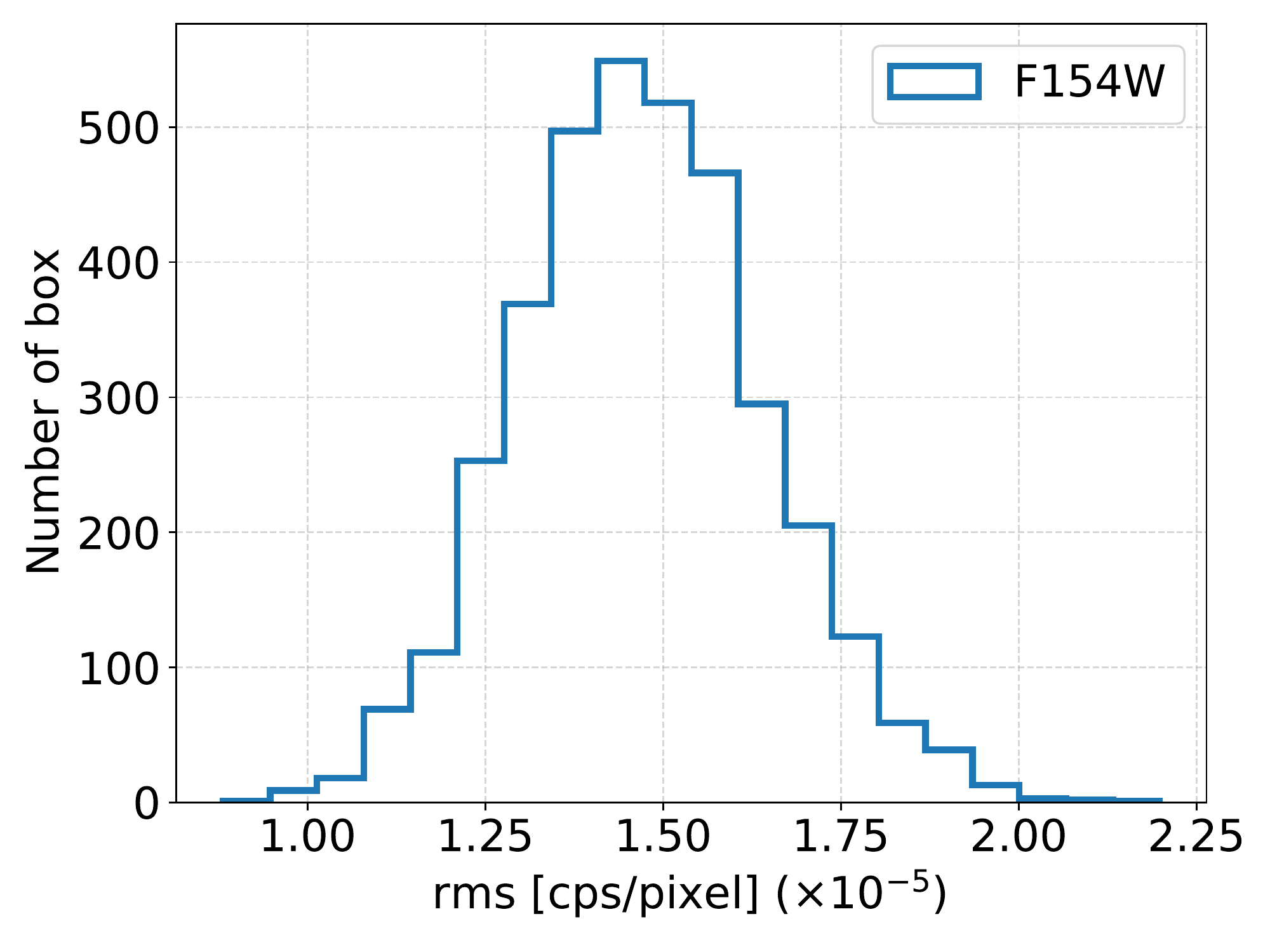}
    \includegraphics[width=2.3in]{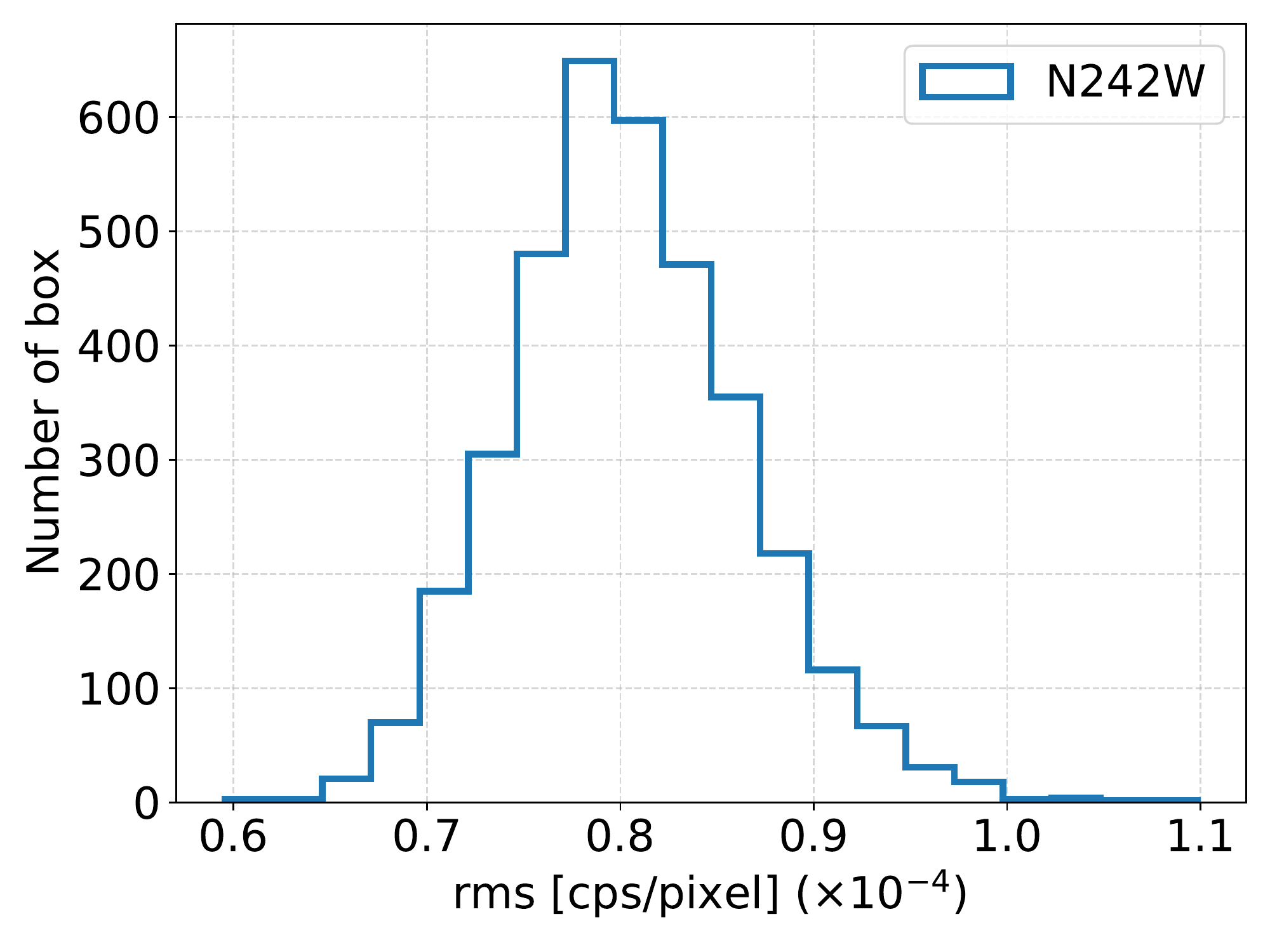}
    \includegraphics[width=2.3in]{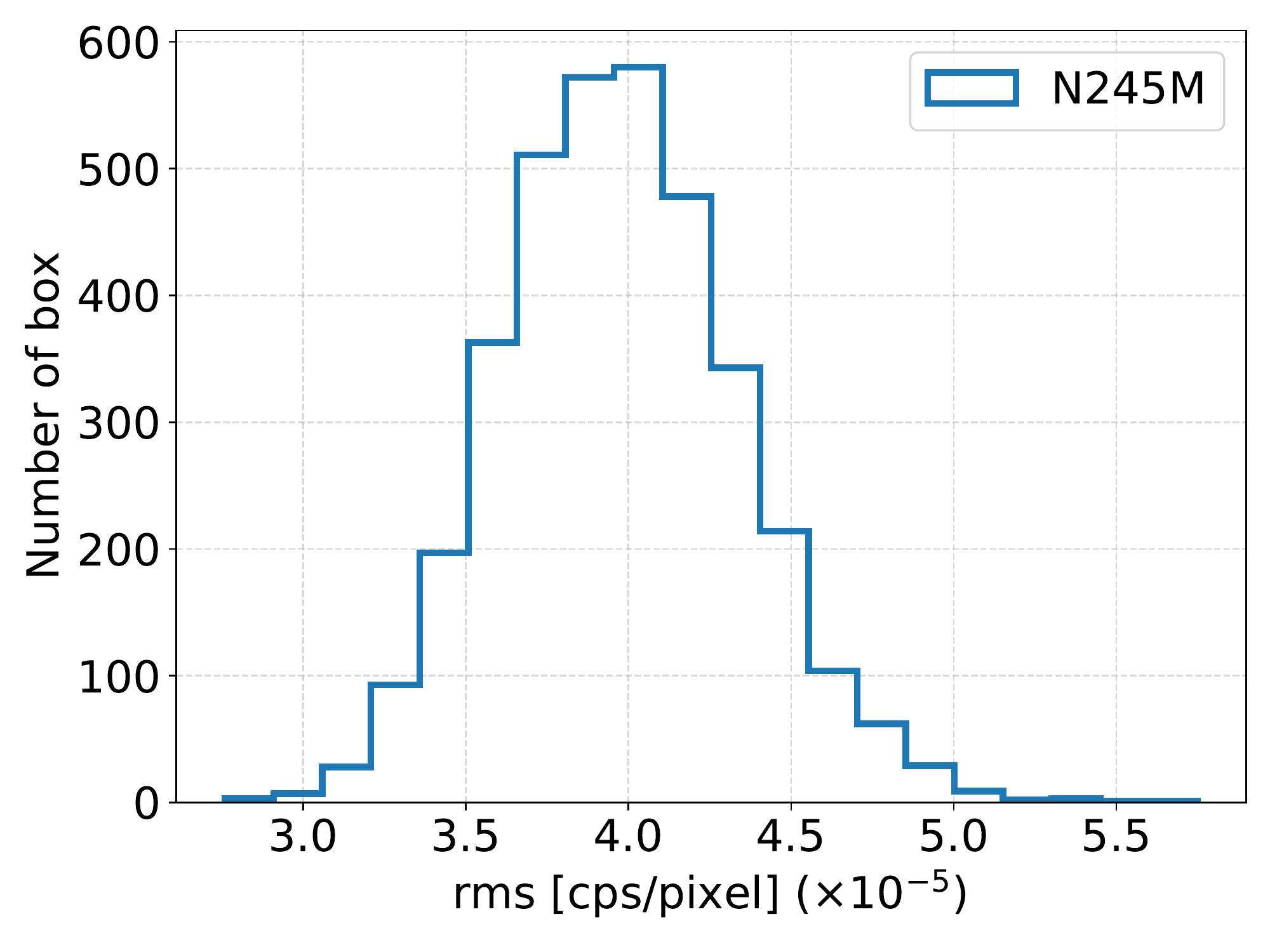}
    \caption{The distribution of the background rms as measured
    within the same 3600 random boxes
    of Fig.~\ref{fig_f154w_bghist} for the same choice of SExtractor parameter values for UVIT F154W (\textit{left}), N242W (\textit{middle}), and N245M (\textit{right}). Here, also, each distribution is one of 100 random iterations.}
    \label{fig_rms_bg}
\end{figure*}

\begin{table}
\centering
\caption{SExtractor parameters used for detection and photometry of sources listed in the FUV and NUV catalog.}
\label{table_se_param}
\begin{tabular}{p{3.5cm}p{4.5cm}}
\hline
Parameter name & Value\\\hline
DETECT\_MINAREA &   10 \\            
DETECT\_THRESH  & 1.25 (F154W), 1.15 (N242W), 1.50 (N245M)  \\ 
FILTER\_NAME &      default.conv \\  
DEBLEND\_NTHRESH &  32  \\          
DEBLEND\_MINCONT & 0.005 \\        
CLEAN\_PARAM     & 1.0 \\          
WEIGHT\_TYPE     & BACKGROUND \\
PHOT\_FLUXFRAC   & 0.5 \\
PHOT\_APERTURES   & 6.71 \\             
PHOT\_AUTOPARAMS  & 2.5, 3.5 \\      
PIXEL\_SCALE      & 0.417 \\                  BACK\_SIZE        & 34 \\            
BACK\_FILTERSIZE  & 3 \\        
\hline
\end{tabular}
\end{table}

\subsection{Background estimation}
\label{s_background_estimation}
The photometric analysis of deep fields requires proper understanding of the sky background as it is crucial to estimate accurate source flux and the signal-to-noise ratio (SNR) of detected objects, especially the fainter ones. Given the dearth of deep field UV observations, particularly in the FUV band, here we explore the UVIT background in detail. To understand the UV background in all the three bands, we used SExtractor v2.25.0 \citep{bertin1996} for initial object detection and masking. We produced a segmentation map for the entire UVIT field in each filter for different detection thresholds before fixing a specific value, which we explain later in this section. To identify sources and produce the segmentation map, we used the default convolution filter. The minimum number of pixels for source identification was fixed to 10, which is equivalent to the area of a circle of diameter $\sim$3.5 pixels (approximating the FWHM of the UVIT PSF). We deblend sources with 32 sub-thresholds and a minimum contrast parameter value of 0.005{\,}.
We used a 34$\times$34 pixel mesh grid with 3$\times$3 filtering to model the background and rms map while producing the segmentation map. We did not clean the detection map since that could result in a segmentation map where some faint sources will be unmasked which may cause a bias in the background measurements. The parameters used in SExtractor for source detection and photometry are listed in Table \ref{table_se_param}.

\begin{table}
\caption{Global mean and rms of the background sky as measured in the three UVIT images.}
\label{table_bg}
\setlength{\tabcolsep}{2pt}
\begin{tabular}{p{1.2cm}p{1.8cm}p{2.1cm}p{1.7cm}p{1.7cm}}
\hline
Filter & Detection & Mean background & RMS (SE) & RMS (RB) \\
 & threshold [$\sigma$] & [cps/pixel] & [cps/pixel] & [cps/pixel]\\
 \multicolumn{1}{c}{(1)} & \multicolumn{1}{c}{(2)} & \multicolumn{1}{c}{(3)} & \multicolumn{1}{c}{(4)} & \multicolumn{1}{c}{(5)}\\\hline
F154W & 1.25 & 6.98$\times$10$^{-6}$ & 1.16$\times$10$^{-5}$ & 1.47$\times$10$^{-5}$\\
N242W & 1.15 & 1.24$\times$10$^{-4}$ & 7.70$\times$10$^{-5}$ & 8.05$\times$10$^{-5}$\\
N245M & 1.50 & 2.51$\times$10$^{-5}$ & 2.94$\times$10$^{-5}$ & 3.97$\times$10$^{-5}$\\
\hline
\end{tabular}
\textbf{Note.} Table columns: (1) Name of the UVIT filter; (2) the adopted value of SExtractor detection threshold; (3) the value of global mean background estimated using random boxes; (4) the value of background rms provided in SExtractor output; (5) the value of mean background rms estimated using random boxes.

\end{table}

\begin{figure}
    \centering
    \includegraphics[width=3.5in]{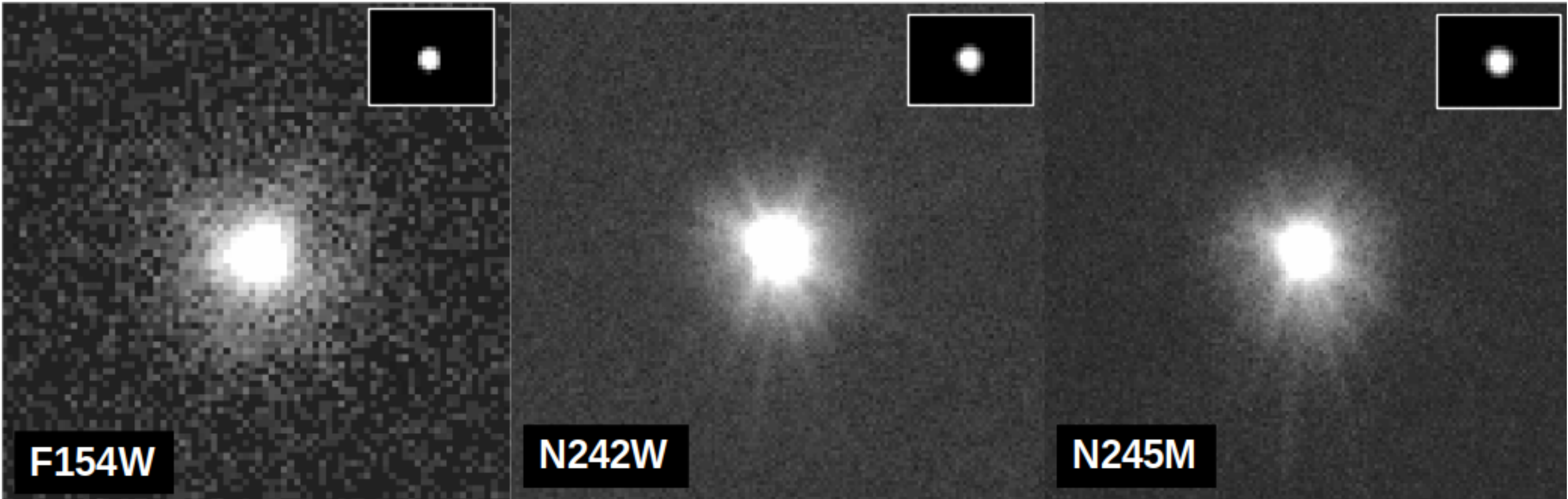}\\
    \caption{The empirical PSFs in three UVIT filters. The NUV PSFs are constructed by stacking the images of 10 stars identified in the UVIT field, whereas we find only 2 candidate stars in the FUV. The panel insets show the core of each PSF.}
    \label{fig_psf_all}
\end{figure}

Following the creation of the segmentation map, we considered nine 6$^{\prime}\times$6$^{\prime}$ square boxes distributed across the UVIT field (Figure \ref{fig_f154w_bg_map}). We placed these boxes such that together they would cover most of the observed field to provide local background values across the field. Using the segmentation map, we placed 400 small random boxes of size 11$\times$11 pixels within each of the defined large boxes. The random boxes were placed such that they contain no pixels from any objects detected in the segmentation map. We generated a total of 3600 random boxes across the field and estimate the average background level in units of cps/pixel for each small box. The procedure was repeated for different values of the SExtractor detection threshold within a range of 1.0 - 3.0\,. For each different threshold, we plot the histogram of cps/pixel of these 3600 random boxes and check its skewness. The detection threshold, for which the histogram had the least skew (i.e., when the difference between the mean and the median is minimum), has been finally selected to detect sources and produce segmentation map. We were able to accomplish more reliable source detection by fixing the threshold in this way. For each of the three bands, we computed the optimal detection threshold following the same approach and the values are listed in Table \ref{table_bg}. We used these same threshold values to produce source catalog for each respective images in \S\ref{s_detection}. For the selected value of detection threshold, we compiled 100 realisations of the random box algorithm and estimate the average value of mean background which we consider as the global mean of the background level. In Figure \ref{fig_f154w_bghist} (\textit{left}), we show one such histogram (among 100 iterations) for the F154W image, produced with the optimal detection threshold of 1.25$\sigma$. To highlight the least skewness of the distribution plotted for the optimal threshold value, we also showed histogram for two other thresholds (which are selected as optimal for the other two filters) in the same figure. The estimated global mean of the F154W band background is 6.98$\times$10$^{-6}$ cps/pixel. We adopt the SExtractor estimate for the rms of 1.16$\times10^{-5}$ cps/pixel as representative for the noise in that background level.

To highlight the variation of the background across the field, we considered the mean of the central box ($B_c$) in Figure \ref{fig_f154w_bg_map} as the reference, and estimated the percentage difference of the local mean ($B_i$) computed for all other boxes. Each of these eight boxes is labelled with the relative difference in background level, 100\%\,$\cdot$\,$(B_i - B_c)/B_c$.
The local mean background within each box was estimated from the distribution of values of the 400 small random boxes in that box.
We find up to $\sim$9\% overall peak-to-peak variation of the FUV background across the 28$^{\prime}$ UVIT field. The variation is even smaller within the portion of the GOODS-N field covered by HST, which is indicated by a red dashed rectangle in Figure~\ref{fig_f154w_bg_map}. 

Using the same method for the N242W and N245M images, the best choices for the detection threshold were found to be 1.15 and 1.50$\sigma$, respectively, with corresponding representative histograms shown in Figure \ref{fig_f154w_bghist} (\textit{middle} and \textit{right} panels). The estimated global mean background levels are (1.24$\pm$0.77)$\times$10$^{-4}$ cps/pixel in N242W and (2.51$\pm$2.94)$\times$10$^{-5}$ cps/pixel in N245M. We also investigated the variation in the background for both the NUV images and find up to $\sim$9\% peak-to-peak variation across the UVIT field. The estimated background values (i.e., the global mean levels listed in Table~\ref{table_bg}) were subtracted from the actual images of the respective band to produce background subtracted images which are used to perform source photometry in \S\ref{s_source_detection}. The variation in the background with respect to the subtracted global mean has a negligible impact on the measured source magnitudes, as its effect is much smaller than the
photometric uncertainties.

To produce background subtracted images, we have not used the SExtractor generated background map in this study. The FUV background map produced by SExtractor contains pixels with negative values which is unrealistic. Because of the low photon events in FUV, the F154W band image contains many pixels with zero value which produce these negative values in the SE background map. Subtracting such a background map will eventually increase source flux in regions that have negative background values. For the same reason, we have not considered local background subtraction while doing source photometry using SE. We noticed a similar problem in NUV N245M band with relatively less pixels with negative values. The background map generated for NUV N242W broadband, which has a higher photon events compared to the other two bands, does not contain pixels with negative values. But to be consistent, we adopted the same approach for background subtraction as explained earlier in this section for all the three images.

\subsection{UVIT Detection Limits}

We estimate the UVIT FUV and NUV 3$\sigma$ detection limits as the magnitudes equivalent to a level of 3 times the rms measured with SExtractor within a circular aperture of radius 1\farcs0 as:
\begin{equation}
\label{e_detection}
    m_{3\sigma}= -2.5\log\left(3\times {\rm rms}\times \sqrt{N_{\rm pix}}\right) + m_{0} \quad ,
\end{equation}
where $N_{\rm pix}$ is the number of pixels within the aperture, and $m_0$ is the
respective zero point magnitude. These estimated detection limits are listed in Table \ref{table_uvit_obs}.
We also used the same random boxes of \S\,\ref{s_background_estimation},
each containing 121 pixels, to check the reliability of the SExtractor background rms in
each band. The distributions of the standard deviations computed in each of the 3600 boxes
have a Gaussian shape and are shown in Figure~\ref{fig_rms_bg}. The mean fit to each
histogram provides the mean rms of the background in the three UV bands as listed in
Table~\ref{table_bg}. The rms estimated using random boxes for the F154W, N242W, and N245M
bands is $\sim$28\%, 5\%, and 35\% higher, respectively, than the corresponding values
reported by SExtractor. The higher values of noise estimated with random boxes might be due to the larger local variations of background on scales of tens of pixels.

\begin{figure}
    \includegraphics[width=3.5in]{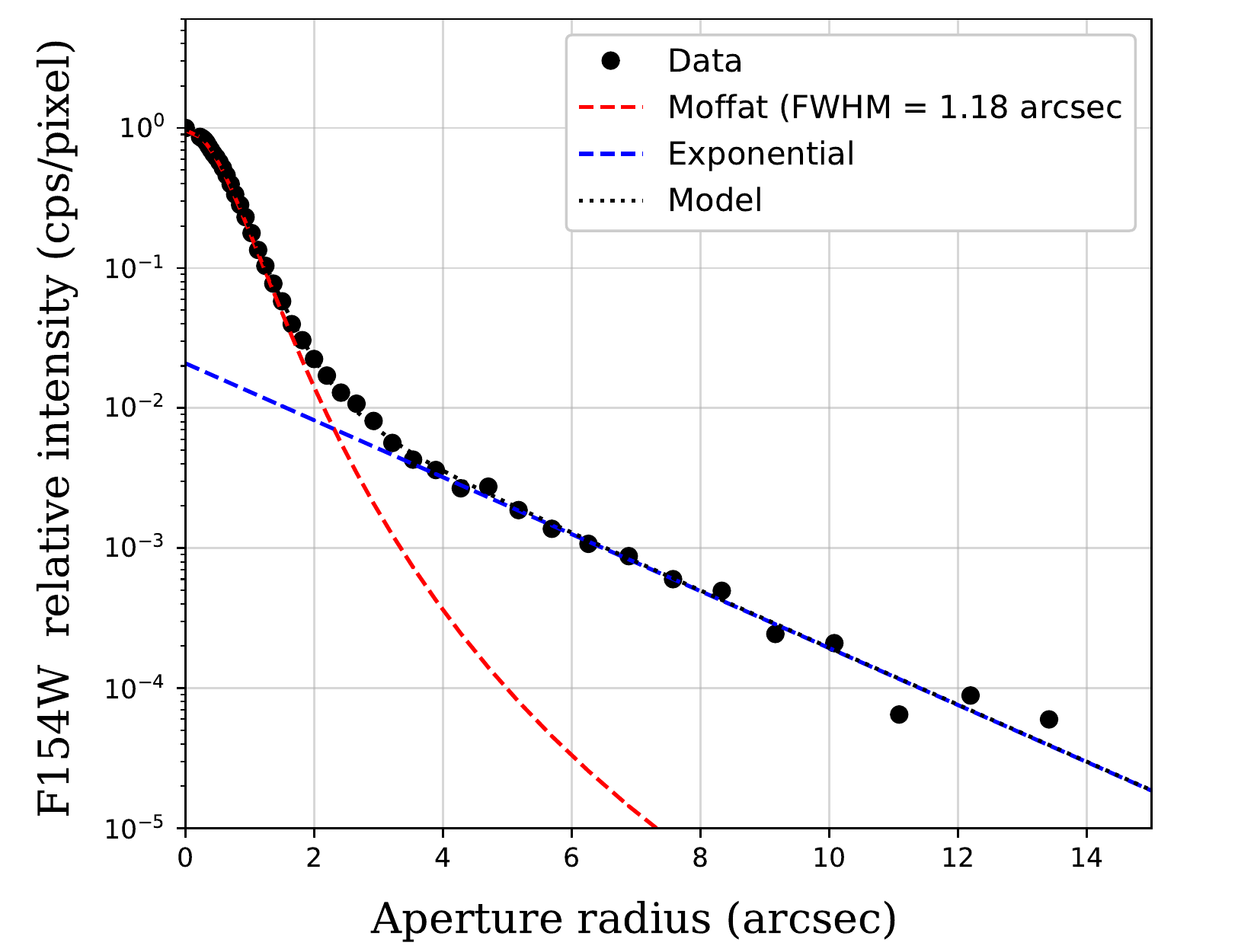}
    \includegraphics[width=3.5in]{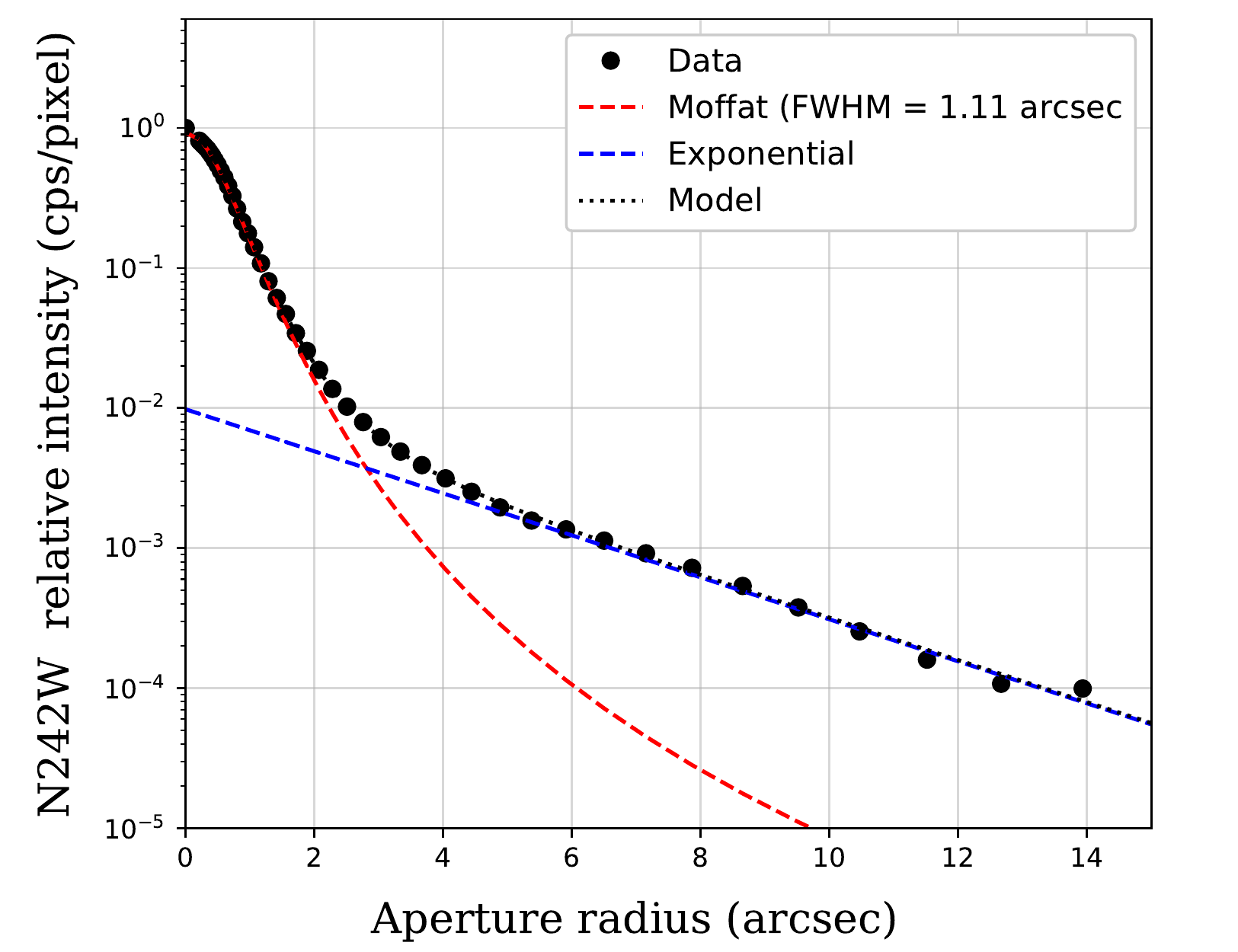}
    \includegraphics[width=3.5in]{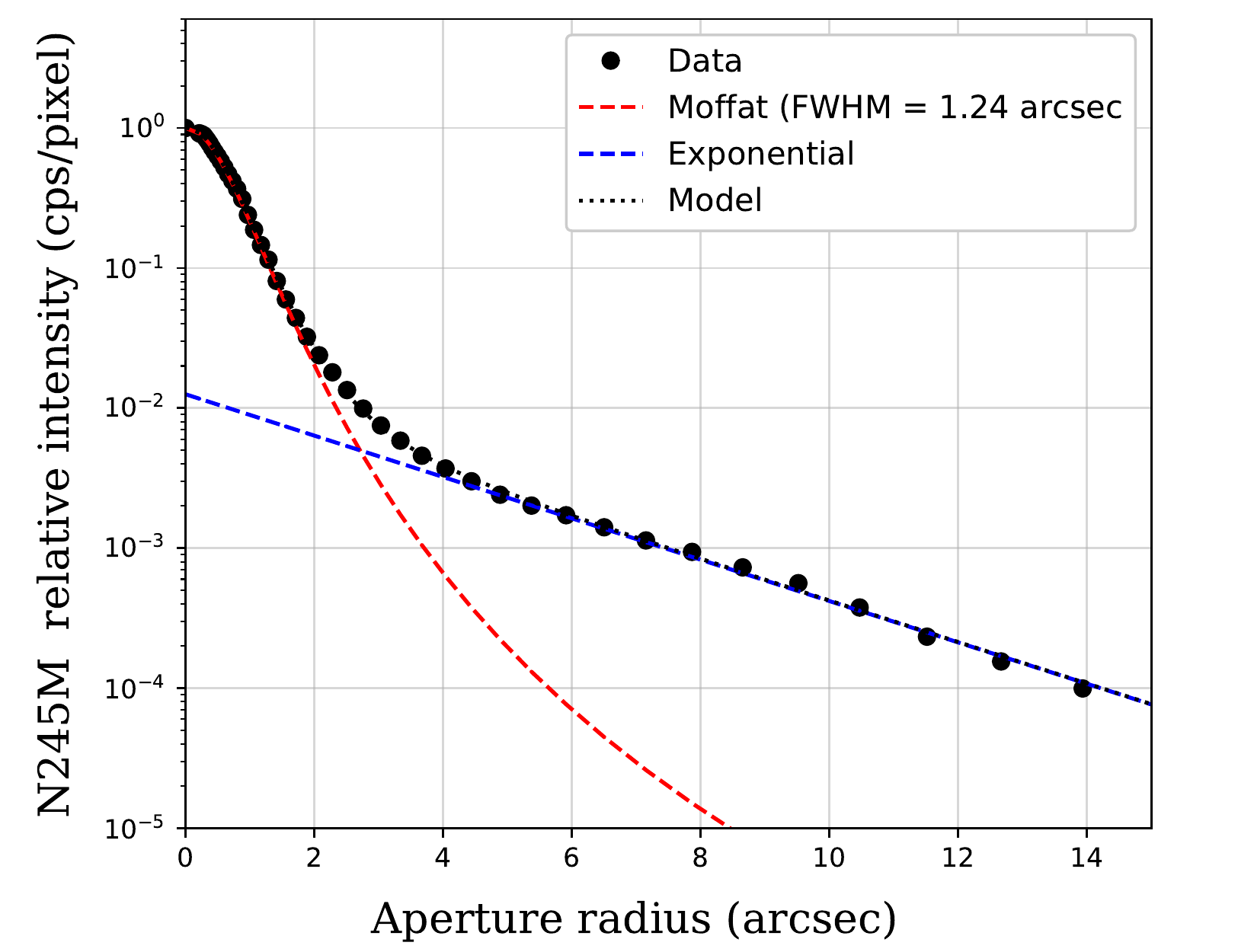} 
    \caption{Radial intensity profiles of the three empirical PSFs shown in Figure~\ref{fig_psf_all} for F154W (\textit{top}), N242W (\textit{middle}), and N245M (\textit{bottom}). The black dots are the isophotal flux measurements. The red dashed curves are Moffat profile fits to the inner portion of each PSF, and the blue dashed lines are exponential fits to the outer wings. The black dotted curve shows the combined model fit to the observed data.}
    \label{fig_psf_f154w}
\end{figure}

\begin{table}
\caption{Parameters estimated from the PSF model fitting}
\label{table_psf}
\begin{tabular}{p{1cm}p{0.8cm}p{1.1cm}p{0.7cm}p{0.5cm}p{1.1cm}p{0.8cm}}
\hline
Filter & FWHM & \centering $I_{m0}$ & $\alpha$ & $\beta$ & \centering $I_{e0}$ & $h_{0}$\\
(1) & (2) & \centering(3) & (4) & (5) & \centering(6) & (7) \\\hline
F154W & 1.18$^{\prime\prime}$ & 0.962248 & 1.18$^{\prime\prime}$ & 3.12 & 0.020850 & 2.14$^{\prime\prime}$\\
N242W & 1.11$^{\prime\prime}$ & 0.926943 & 0.98$^{\prime\prime}$ & 2.49 & 0.009804 & 2.90$^{\prime\prime}$\\
N245M & 1.24$^{\prime\prime}$ & 0.999933 & 1.20$^{\prime\prime}$ & 2.93 & 0.012547 & 2.94$^{\prime\prime}$\\
\hline
\end{tabular}
\textbf{Note.} Table columns: (1) name of the UVIT filter; (2) the PSF FWHM calculated from the fitted Moffat parameters; (3)--(5) fitted values of the three parameters ($I_{m0}$, $\alpha$, $\beta$) associated with the Moffat function portion of Eqn.~\ref{e_psf}; (6)--(7) fitted values of the two parameters ($I_{e0}$, $h_{0}$) associated with the exponential portion of Eqn.~\ref{e_psf}.
\end{table}

\subsection{Point Spread Function profiles}

We used the GAIA source catalog \citep{gaia2021} to identify bright stars within the observed UVIT field. In order to select a robust sample for PSF modelling, we only used bright stars with F154W/N242W magnitude $<20.5$.
Following a careful inspection of the light profile of each star, we found 2 stars in the FUV and 10 stars in the NUV bands to be suitable. We stacked the stellar images to produce empirical PSFs in each band, after aligning their centroids and removing pixels with possible contamination from neighboring objects. The resulting empirical PSFs are shown in Figure~\ref{fig_psf_all}.

\begin{table}
\centering
\caption{Percentage of light enclosed within different aperture radius for each PSF as shown in Figure \ref{fig_cog}}
\label{table_cog}
\begin{tabular}{p{2.5cm}p{1cm}p{1cm}p{1cm}}
\hline
Aperture & F154W & N242W & N245W\\
radius (arcsec) & \% & \% & \% \\\hline

0.42 & 19.2 & 16.0 & 15.6 \\
0.83 & 48.3 & 45.5 & 42.7\\
1.25 & 66.6 & 63.6 & 61.9\\
1.40 & 69.7 & 66.8 & 65.3\\
1.67 & 75.4 & 72.5 & 71.4\\
2.09 & 80.5 & 77.9 & 77.2\\
2.50 & 83.7 & 81.2 & 80.8\\
2.92 & 86.2 & 83.7 & 83.4\\
3.34 & 88.1 & 85.6 & 85.2\\
3.75 & 89.7 & 87.2 & 86.8\\
4.17 & 90.9 & 88.5 & 88.1\\
5.42 & 94.1 & 91.6 & 91.1\\
6.67 & 96.5 & 94.1 & 93.5\\
7.92 & 98.0 & 96.0 & 95.5\\
9.17 & 98.9 & 97.6 & 97.1\\
10.43 & 99.5 & 98.6 & 98.3\\
11.68 & 99.8 & 99.2 & 99.1\\
12.93 & 99.9 & 99.7 & 99.7\\
13.97 & 100.0 & 100.0 & 100.0\\
\hline
\end{tabular}
\end{table}

\begin{figure}
    \centering
    \includegraphics[width=0.48\textwidth]{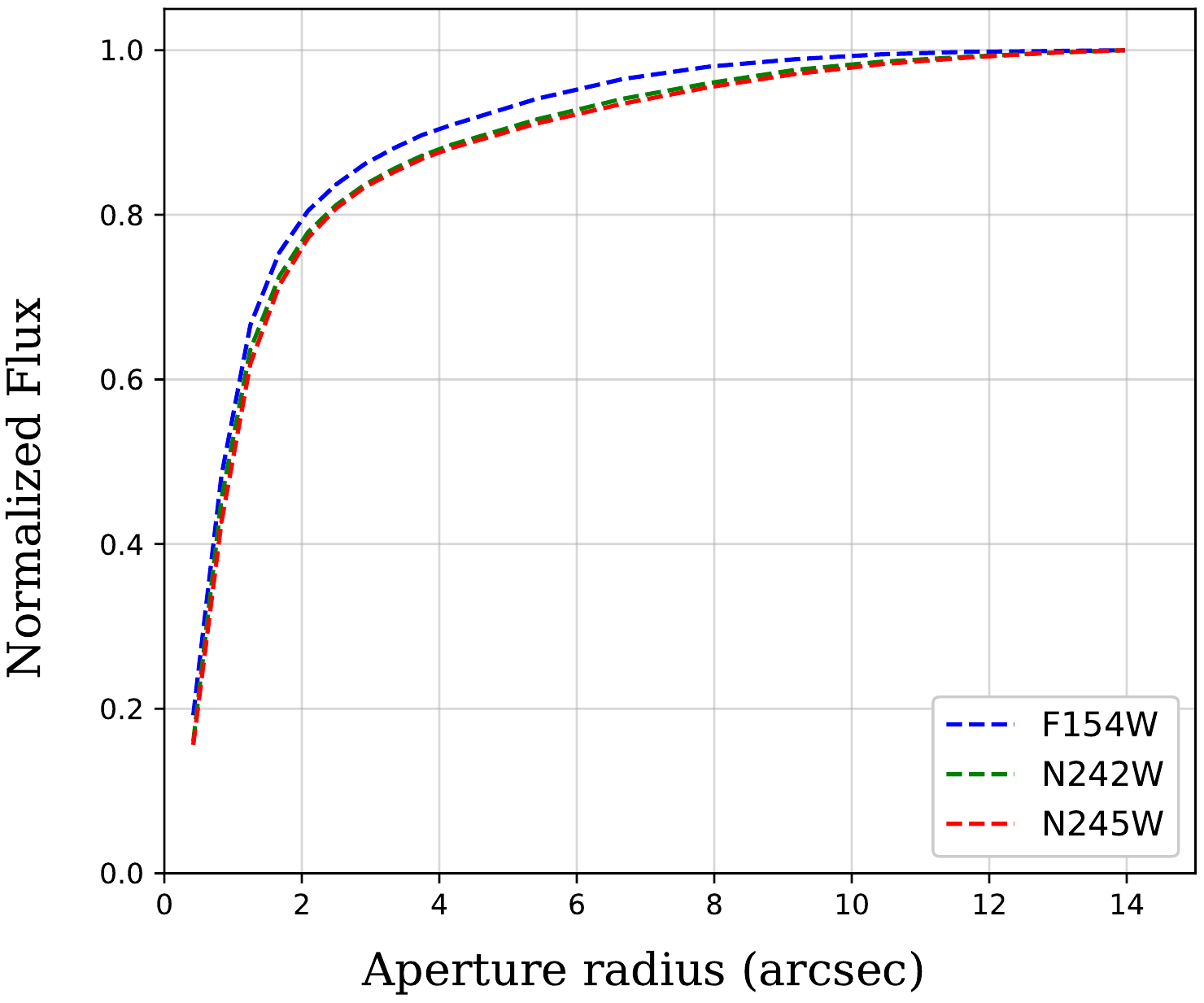} 
    \caption{Curve of growth profiles of the PSFs in the F154W (\textit{shown in blue}), N242W (\textit{green}), and N245M (\textit{red}) bands. The figure shows the percentage of the total flux (normalised) contained within apertures with increasing radii.}
    \label{fig_cog}
\end{figure}

\begin{figure*}
    \centering
    \includegraphics[width=3.2in]{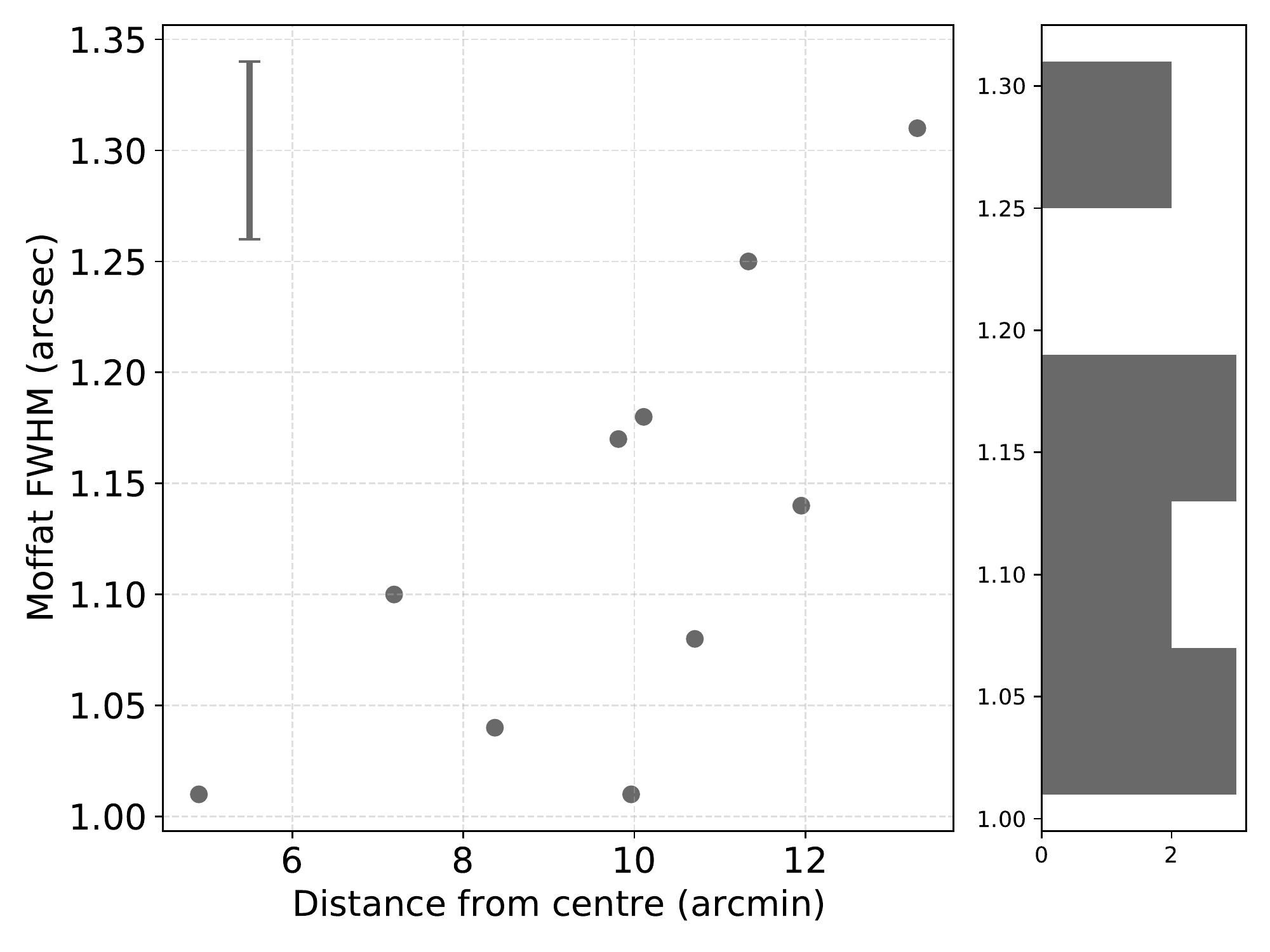} 
    \includegraphics[width=3.2in]{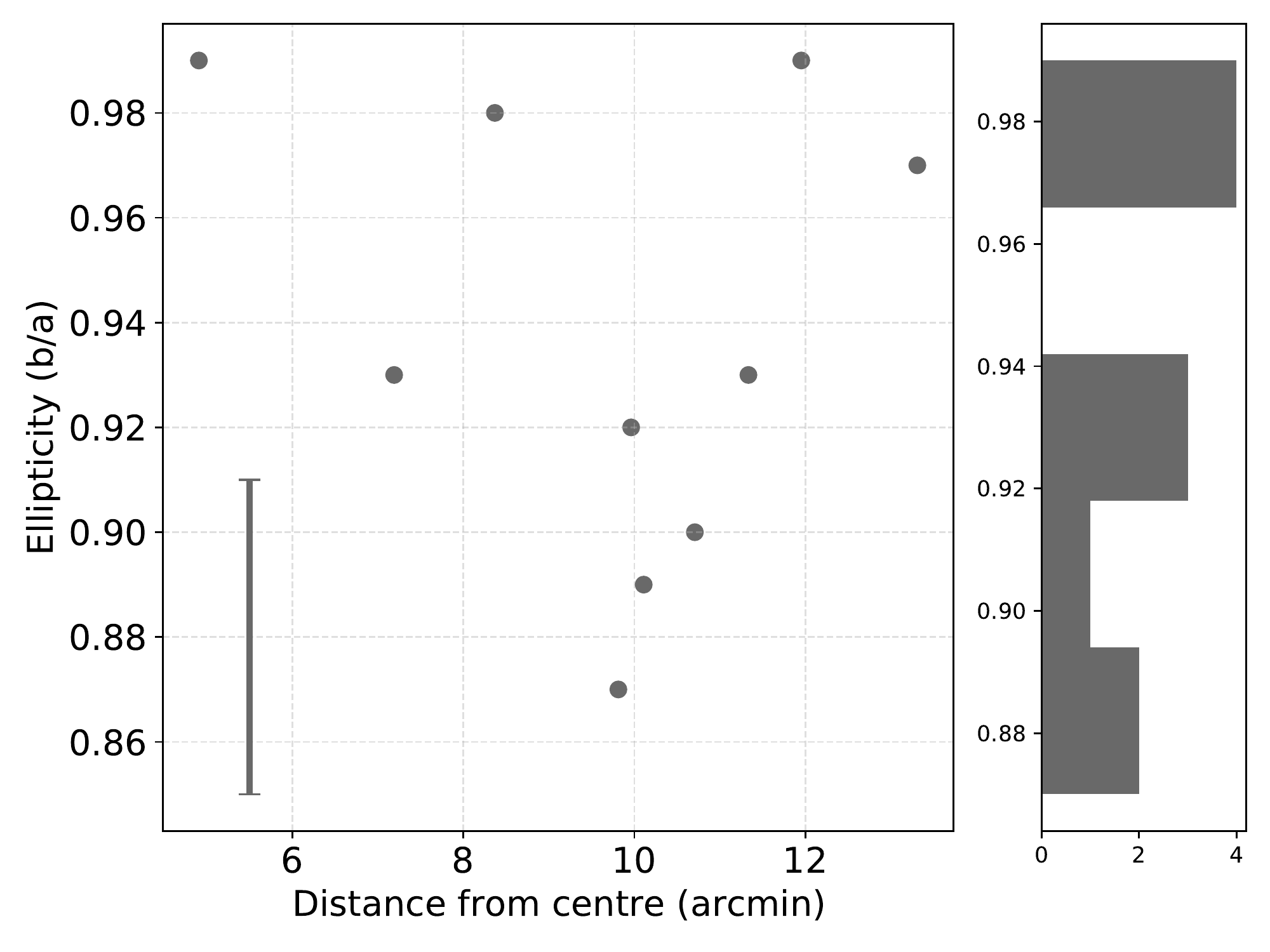} 

    \caption{Left: The FWHM of 10 selected stars in N242W bands are plotted with their distance from the field centre. The adjacent histogram shows the distribution of measured FWHM of the selected stars. The errorbar plotted on the top-left shows the typical error in the estimated FWHM. Right: The ellipticity (b/a) of 10 selected stars in N242W bands are plotted with their distance from the field centre. The adjacent histogram shows the distribution of measured ellipticity of the selected stars. The errorbar plotted on the bottom-left shows typical error in the estimated ellipticity.}
    \label{fig_psf_fwhm_eli}
\end{figure*}

We used the \textit{ellipse} procedure (Jedrzejewski 1987) within IRAF \citep{tody1986} to fit the isophotal flux density as a function of semi-major axis for each PSF image, and produced the radial intensity profiles shown in Figure~\ref{fig_psf_f154w}. The observed PSF intensity profile of each filter was fitted with a combination of a Moffat profile \citep{moffat1969} and an exponential function, as given by:
\begin{equation}
\label{e_psf}
    I(r)= I_{m0} \left[ 1+\left(\frac{r}{\alpha}\right)^2 \right]^{-\beta}
       + I_{e0}\, e^{-r/{h_0}} \quad .
\end{equation}
The Moffat function (the first term in the equation) fits the core
portion of the PSF, while the exponential function (second term) is used to fit its wing. The value of each parameter, $I_{m0}$, $I_{eo}$, $\alpha$, $\beta$, and $h_0$, was fit and is listed in Table~\ref{table_psf}. The best PSF model fits for the F154W, N242W, and N245M filters are also shown in Figure~\ref{fig_psf_f154w}.
Estimating the FWHM of each PSF from its Moffat core as 2$\alpha\sqrt{2^{1/\beta}-1}$, we find PSF FWHM values of 1\farcs18, 1\farcs11, and 1\farcs24 for F154W, N242W, and N245M, respectively.

The wing of each PSF was fit up to a radius of $\sim$14\farcs0 ($\sim$34 pixels), since the flux density around individual stars becomes similar to the estimated background mean at this radius. We plotted the curve of growth (COG) of each PSF up to this radius in Figure \ref{fig_cog}. The COG profiles convey the percentage of light contained up to a particular aperture radius for each PSF. This quantity is important for the accurate flux estimation of galaxies, particularly the fainter ones. Modelling the PSF wing is crucial for determining the amount of light scattered outside the core of the PSF. In Table \ref{table_cog}, we indicate the percentage of light enclosed within a particular aperture radius of each PSF. The table provides the necessary information to decide the aperture size for photometry along with the growth curve correction factor. Our analysis shows that around 80\% of the total light is contained within $\sim$\,2\farcs1\,--\,2\farcs4 aperture radii in the three bands.  

\begin{figure}
    \centering
    \includegraphics[width=3.5in]{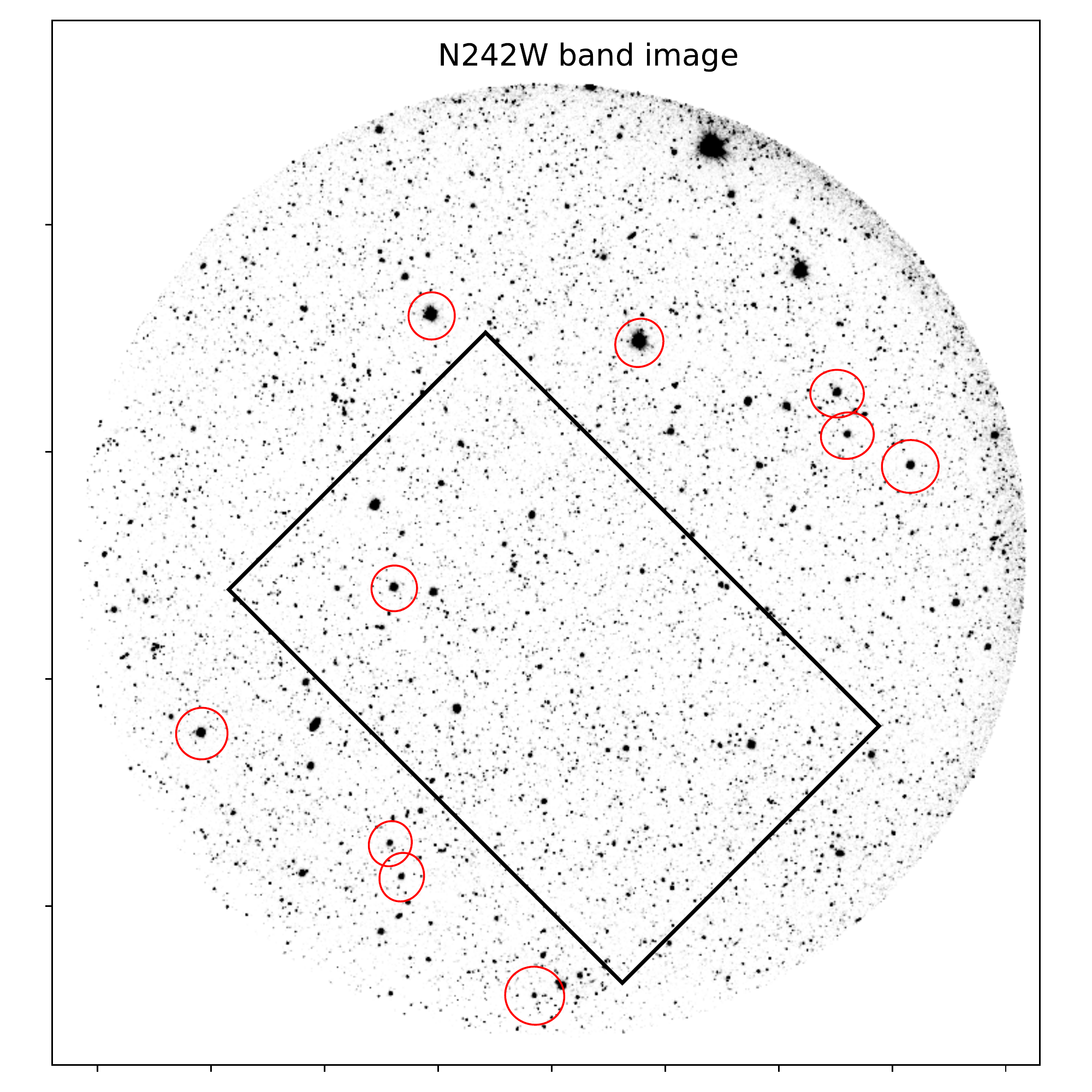} 
    \caption{The stars selected for constructing the empirical PSF in NUV are marked with red ellipse. The size of each ellipse is scaled up by multiplying a constant factor with the FWHM of that star. The ellipticity shows the actual measured value. The region marked with black rectangle shows the HST coverage.}
    \label{fig_psf_stars}
\end{figure}

Because we had a good number of candidate PSF stars in the NUV band, we utilised each of them to understand the variation of PSF FWHM and ellipticity across the field. To achieve this, we took the observed light profile of each of the 10 selected stars and fit them to a Moffat profile as explained above in order to estimate their FWHM. In Figure \ref{fig_psf_fwhm_eli} (left), we show the distribution of the FWHM values as a function of the star's distance from the centre of the field. The FWHM of the stars is seen to vary between $\sim$ 1\farcs0\,--\,1\farcs4, and show a trend of increasing size with increasing distance from the field centre. Since the selected stars have brighter magnitudes, we tested the effect of saturation on the estimated FWHM. Using the equation provided in \citet{tandon2020}, we determined the saturation corrected radial intensity profile for each star and re-calculated the FWHM. We find only $\sim$1\% (i.e., $\sim 0.01^{\prime\prime}$) change in the FWHM value for two stars brighter than 17.0 magnitude. The FWHM of the remaining 8 stars, which have magnitude between 18.0 - 20.5, remains the same. The 2D isophotal contours of these candidate stars were also not always circular. We estimated the ellipticity of the selected stars using the measurements obtained from IRAF. Using the isophote that contained 80\% of the total light of the star as the reference, we estimated the ellipticity b/a (i.e., the ratio of semi-minor to semi-major axis) at that radius. The derived ellipticity shows a range between $\sim$\,0.87\,--\,0.99 (Figure \ref{fig_psf_fwhm_eli} (right)). The variation of FWHM and ellipticity of the selected stars in the N242W band image is shown in Figure \ref{fig_psf_stars}, marking each star with an ellipse. The size of the ellipse is scaled by a constant factor with the estimated FWHM of that star and the ellipticity reflects the actual measured value. The figure demonstrates the overall variation of the PSF across the observed UVIT field. We could not perform the same such exercise in the F154W band due to few available candidate stars in the FUV.

\begin{figure*}
    \centering
    \includegraphics[width=6.5in]{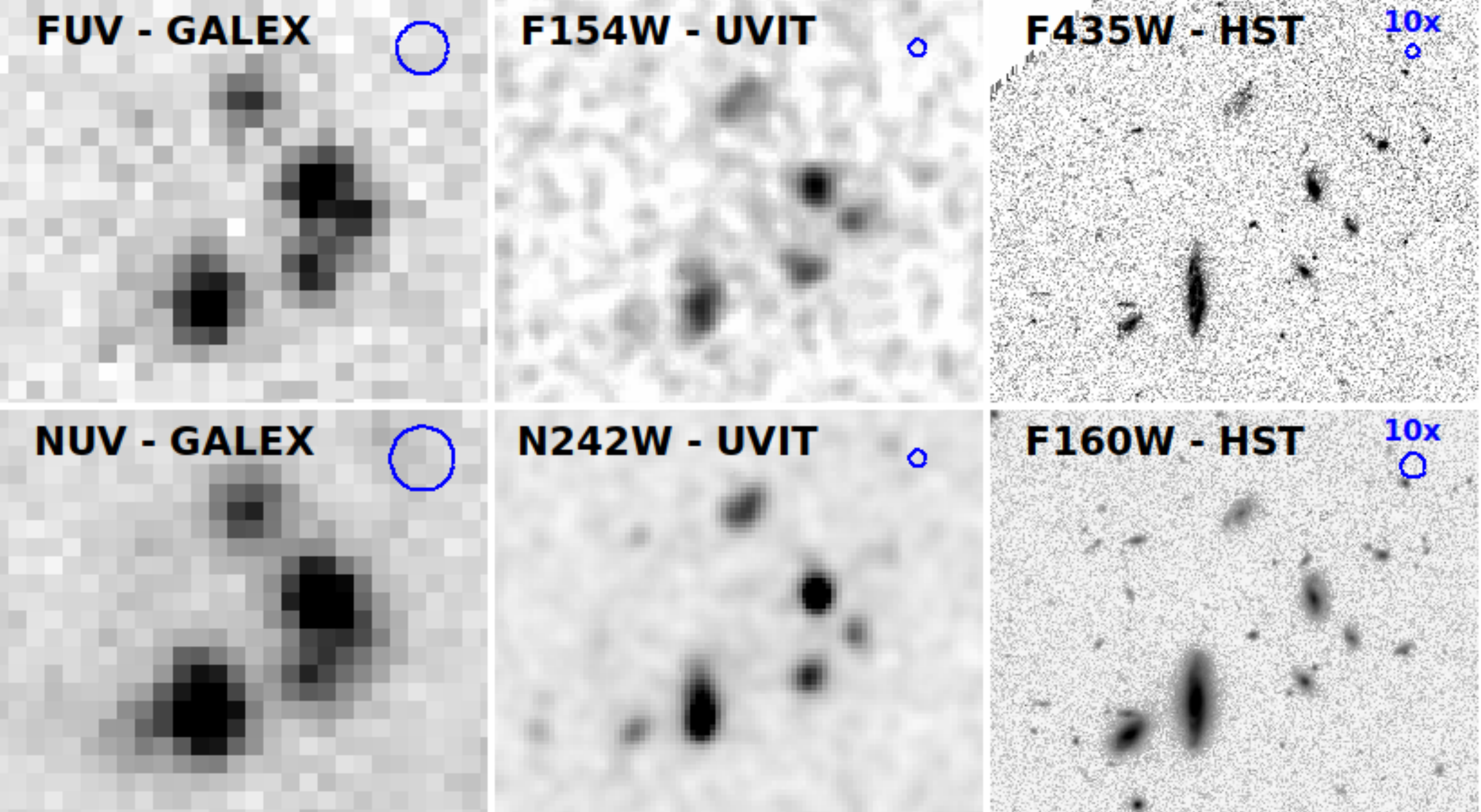} 
    \caption{A selected region in the GOODS-N field is shown as observed in different filters by GALEX (FUV, NUV), AstroSat/UVIT (F154W, N242W), and HST (F435W, F160W). The figure highlights the difference in angular resolution for three different telescopes. The size of the PSFs is indicated by blue circles in each panel. Both the HST PSF sizes are shown with 10$\times$ enhanced size.}
    \label{fig_uvit_galex}
\end{figure*}

\section{Source Detection and Photometry}
\label{s_source_detection}
\subsection{Detection}
\label{s_detection}
To identify sources in each image, we use SExtractor with the same adopted input parameters while creating the segmentation maps described in \S\ref{s_background_estimation}. The source catalog we aim to provide here is primarily motivated for the identification of LyC leaking galaxies using either of the FUV or NUV bands. This demands robust detection and flux measurement in individual images. As there can be genuine structural differences in a galaxy between FUV and NUV filters, the photometric centroid could differ for the same galaxy in either of these filters. Thus, the conventional approach of creating a detection image, i.e., combining images from both the bands, for identifying sources may result in less accurate flux measurements. So we produce two separate catalogs - one for FUV and other for the NUV image considering detection in each respective band. This method improves the placement of the aperture centred on the photometric centroid of the object in each individual band, ensuring correct flux measurements. However, an accurate color measurement would require using same aperture around a common centroid. To make our measurement more useful, in each catalog we performed photometry on all the three bands using the detection from the respective catalog. Therefore, addition to the flux measurement in F154W band, the FUV catalog also provides magnitude in both the NUV bands (N242W, N245M) for the FUV-detected sources and the vice versa. As the UVIT field coverage extends beyond the HST-covered GOODS-north region, in both of these catalogs we have not used positional prior from the high-resolution HST images for photometry in the UVIT images. Although, we provide a separate UV photometric catalog using the positional prior of sources from the HST-CANDELS catalog of the GOODS-north field \citep{barro2019}. We discuss this in Section \S\ref{s_candels}.

The UV sources common to both FUV and NUV images can be found by cross-matching both the catalogs. We employed both the actual and background subtracted images (from Section \ref{s_background_estimation}) in dual mode to identify sources and measure their flux. The actual images in F154W and N242W, weighted by the background, were used to identify objects in FUV and NUV respectively, whereas photometry was performed on the corresponding background subtracted images. We cleaned the list of identified objects by fixing the cleaning parameter to 1\,. This helped to exclude some spurious faint detections around the bright objects. The photometry in N245M was carried out using a different approach. As the N245M passband falls within the N242W broad band (Figure \ref{fig_filters}), we used the N242W band image as the detection image and measured the flux from the background-subtracted N245M band image.

The efficiency of source detection is sensitive to the angular resolution of a telescope. To demonstrate the capability of UVIT, we show images of a particular region of the GOODS-N field as observed with GALEX, UVIT, and HST in Figure \ref{fig_uvit_galex}. The figure highlights the superior resolution of UVIT, which allows us to resolve sources in crowded regions better than GALEX, and provides more secure source identifications when cross-matching sources with the extant HST imagery and catalogs at longer wavelengths.

\begin{table}
\caption{Aperture correction magnitudes of the UVIT and HST (for which flux measurements are provided in CANDELS catalog by \citet{barro2019}) filters.}
\label{table_aper_corr}
\begin{tabular}{p{2cm}p{2.5cm}p{2.5cm}}
\hline
Filter & \multicolumn{2}{c}{Aperture correction value (mag)}\\\hline
 & r = 0\farcs7 & r = 1\farcs4\\\hline
F154W & 1.02 & 0.39 \\
N242W & 1.10 & 0.44\\
N245M & 1.17 & 0.46\\\hline
F435W & 0.14 & 0.09\\
F606W & 0.11 & 0.07\\
F775W & 0.09 & 0.05\\
F814W & 0.08 & 0.05\\
F850LP & 0.10 & 0.06\\
F105W & 0.10 & 0.03\\
F125W & 0.11 & 0.04\\
F160W & 0.12 & 0.05\\\hline
\end{tabular}
\textbf{Note.} Table columns: (1) name of the filter; (2) aperture correction magnitude for an aperture of radius 0\farcs7\,; (3) aperture correction magnitude for an aperture of radius 1\farcs4\,.
\end{table}

\begin{table*}
\centering
\caption{Content of the NUV catalog.$^{*}$}
\label{table_cat_content}
\begin{tabular}{p{3.5cm}p{13.5cm}}
\hline
Column name & Description\\\hline
ID & Identification number in the catalog\\
X\_image & X pixel coordinate of the centroid\\
Y\_image & Y pixel coordinate of the centroid\\
RA & RA (J2000) of the object's centroid\\
DEC & Dec (J2000) of the object's centroid\\
N242W\_kron\_m & N242W magnitude estimated with kron aperture\\
N242W\_kron\_merr & Error in N242W\_kron\_m\\
N242W\_m & N242W magnitude estimated with a circular aperture of radius 1.4$^{\prime\prime}$ (including aperture correction)\\
N242W\_merr & Error in N242W\_m magnitude\\
N245M\_kron\_m & N245M magnitude estimated with kron aperture\\
N245M\_kron\_merr & Error in N245M\_kron\_m\\
N245M\_m & N245M magnitude estimated with a circular aperture of radius 1.4$^{\prime\prime}$ (including aperture correction)\\
N245M\_merr & Error in N245M\_m magnitude\\

F154W\_kron\_m & F154W magnitude estimated with kron aperture\\
F154W\_kron\_merr & Error in F154W\_kron\_m\\
F154W\_m & F154W magnitude estimated with a circular aperture of radius 1.4$^{\prime\prime}$ (including aperture correction)\\
F154W\_merr & Error in F154W\_m magnitude\\

flux\_radius & Radius that encloses 50\% of the object's total light in N242W\\
kron\_radius & Kron aperture radius in N242W\\
a\_image & Semi-major axis (A\_IMAGE in SExtractor)\\
a\_image\_err & Error in a\_image\\
b\_image & Semi-minor axis (B\_IMAGE in SExtractor)\\
b\_image\_err & Error in b\_image\\
PA & position angle of the source\\
flag1 & Object within the HST coverage (flag1 = 1), object outside HST coverage (flag1 = 0)\\
flag2 & Object is detected in FUV image also (flag2 = 1), object is detected only in NUV (flag2 = 0)\\
\hline
\end{tabular}
$^*${A similar table is provided for the FUV catalog as well}
\end{table*}

\begin{figure}
    \centering
    \includegraphics[width=3.3in]{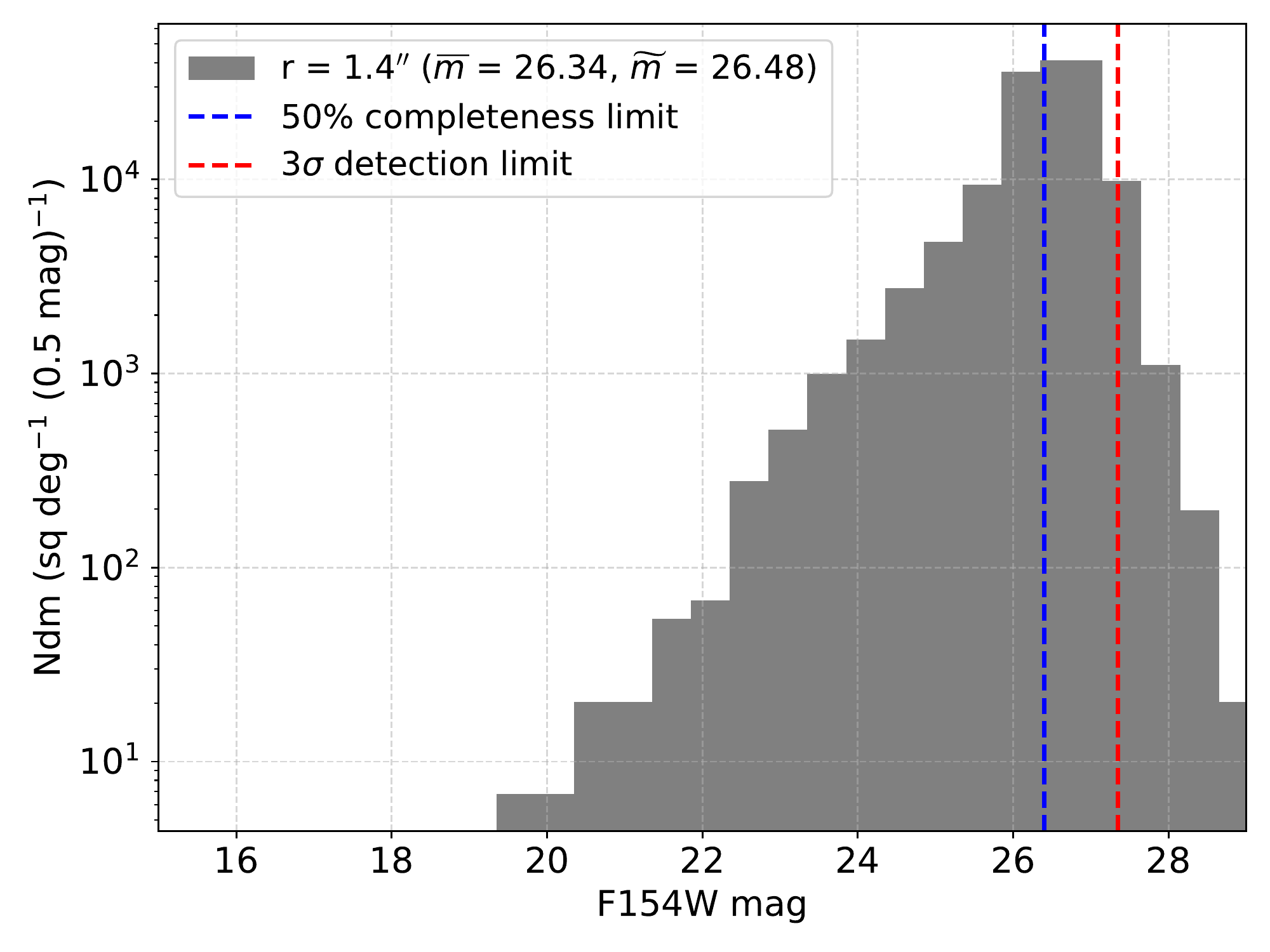}
    \includegraphics[width=3.3in]{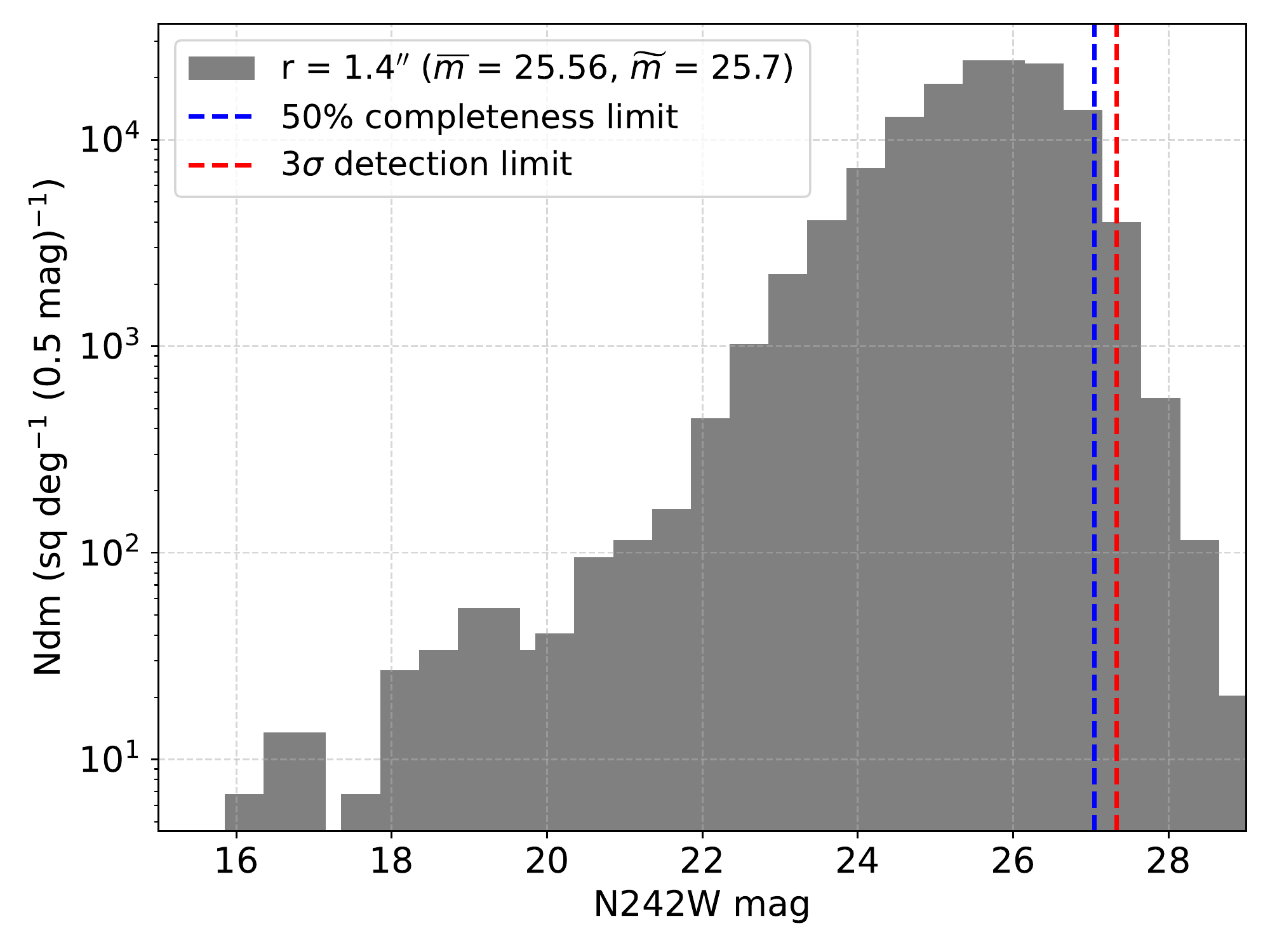}
    \includegraphics[width=3.3in]{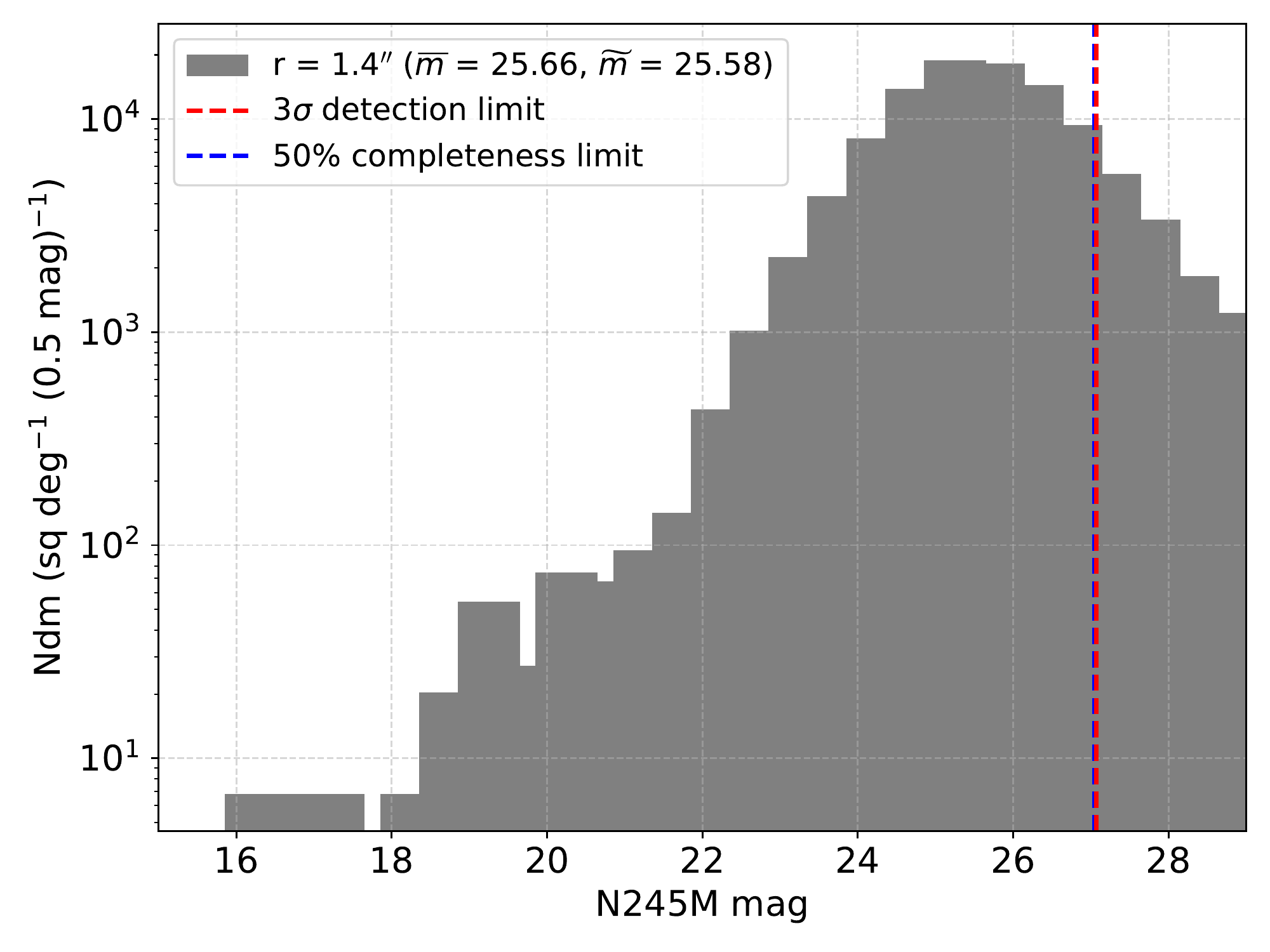}
    \caption{Differential number count (number/deg$^2$/0.5\,mag) histograms
    versus AB magnitude of sources detected within 13$^{\prime}$ from the UVIT field centre (i.e., within the black dashed circle shown in Figure \ref{fig_f154w_bg_map}) for F154W (\textit{top}), N242W (\textit{middle}), and N245M (\textit{bottom}). All source photometry used circular apertures with a radius of 1\farcs4, and was corrected using the aperture correction derived from the PSF for that filter. The mean ($\overline{m}$) and median ($\widetilde{m}$) values of each distribution are noted in the figure. The red and blue dashed lines mark the 3$\sigma$ detection and 50\% completeness limit, respectively. The photometry in the N245M band was performed in dual mode using the N242W band image as the detection image, whereas the F154W and N242W source detections and photometry were performed in those images individually.}
    \label{fig_f154w_mag_hist}
\end{figure}

\subsection{Photometry and Catalog content}

Source photometry was performed using SExtractor with a fixed circular aperture. We inspected the growth curve profile of PSFs, the average size of detected objects, and the retrieved flux of artificially injected point sources (discussed in \S\ref{sect:completeness}) to choose an optimal aperture radius of 1\farcs4 for photometry. This aperture encloses $\sim$69.7\%, 66.8\%, and 65.3\% of the total light for PSFs in F154W, N242W, and N245M filters, respectively. We applied an aperture correction by incorporating the amount of flux contained outside a radius of 1\farcs4\,. The measured values of aperture correction are 0.39 mag, 0.44 mag, and 0.46 mag in F154W, N242W, and N245M bands, respectively (Table \ref{table_aper_corr}). We note that a larger aperture adds more background noise to the source flux, which can significantly affect the photometry of fainter sources. In Figure \ref{fig_f154w_mag_hist}, we show the differential number density of identified sources located within 13.0$^{\prime}$ (i.e., within the black dashed circle shown in Figure \ref{fig_f154w_bg_map} which covers $\sim$ 531 square arcmin area) from the UVIT field centre for all three bands. To avoid including spurious source detection and less accurate photometric measurements, we excluded sources identified beyond 13$^{\prime}$ radius, which corresponds to a circular annulus of width $\sim$ 1$^{\prime}$ from the edge of the detector. 

The magnitudes reported in our FUV and NUV source catalogs (i.e., F154W\_m, N242W\_m, N245M\_m) were measured using 1\farcs4 apertures and corrected using the aperture correction derived from our empirical PSFs and COG analysis. For magnitudes brighter than m\,$\leq$\,26, the source number density in NUV is clearly higher compared to the FUV. \citet{windhorst2011} reported similar counts for the differential source number density in NUV for the GOODS-south field observed with F225W and F275W HST filters. The galaxy count slope in N242W band, estimated between the magnitude range 18 - 25 mag, has a value of 0.44 dex mag$^{-1}$, which matches with the same estimated in HST F225W band by \citet{windhorst2011} for GOODS-south field. For the F154W band, the slope is measured as 0.57 dex mag$^{-1}$ for magnitude range 19 - 25 mag. The approach we adopted to measure flux (i.e., fixed circular aperture photometry) fits well for point-like objects, whereas in the case of extended galaxies of larger angular size our flux measurement  with 1.4$^{\prime\prime}$ circular aperture is not robust. For extended objects, the aperture correction factor estimated using the curve of growth of PSFs can not ideally account for the amount of flux contained outside the aperture of radius 1.4$^{\prime\prime}$. Therefore, we also provide source magnitudes measured using the Kron aperture (i.e., kron magnitude), which yields more robust flux values for extended objects.

We identified a total of 16,001 and 16,761 sources respectively in FUV and NUV within 13$^{\prime}$ radius of the UVIT field. Our catalog contains 6839, 16171, and 13577 sources brighter than the 50\% completeness limit (estimated in \S\ref{sect:completeness}) in F154W, N242W, and N245M bands, respectively. The number of sources detected within the HST covered region (red dashed rectangle in Figure \ref{fig_f154w_bg_map}) is 6,082 (FUV) and 6,292 (NUV) with 2486 and 6038 brighter than the 50\% completeness limit in respective bands. The rectangle encloses $\sim$ 17$^{\prime} \times$11$^{\prime}$ area in the sky centred on (RA, Dec)$_{\rm J2000}$ = (12:36:55.8, +62:14:23.3). In each of the catalogs, objects located inside the HST coverage is flagged as $flag1 = 1$, whereas $flag1 = 0$ signifies sources outside the red rectangle. Sources with $flag2 = 1$ signify detection in both the FUV and NUV bands (for a cross-match radius of 1\farcs4), whereas $flag2 = 0$ means detection only in that corresponding band.

Each histogram in Figure \ref{fig_f154w_mag_hist}, shows a significant drop in the source number beyond the 3$\sigma$ detection limit. Among the two NUV bands, we found sources fainter than 3$\sigma$ detection limit are more in N245M than N242W. This is due to the method of photometry adopted for the N245M filter. As for the N245M band photometry, we used the detection image from the NUV broadband (N242W), sources fainter in the NUV medium band (N245M) simply populate the fainter side of the N245M histogram. We also provide the measured value of Kron radius, size of the semi-major and semi-minor axes (including error) of the kron aperture and its position angle in the catalog. These quantities convey the structural properties of the detected sources. We estimated the flux radius, which is defined as the aperture that contains 50\% of the object's total light, for the identified sources. The list of measured quantities are shown in Table \ref{table_cat_content}. The catalogs are made available in electronic format. Both of our catalogs contain all the sources detected in each respective bands, but the photometry of the objects fainter than 50\% completeness limit is less reliable due to the background contribution. We caution the reader to be more careful while considering the measurements for such faint sources.

\begin{figure*}
    \centering
    \includegraphics[width=0.33\textwidth]{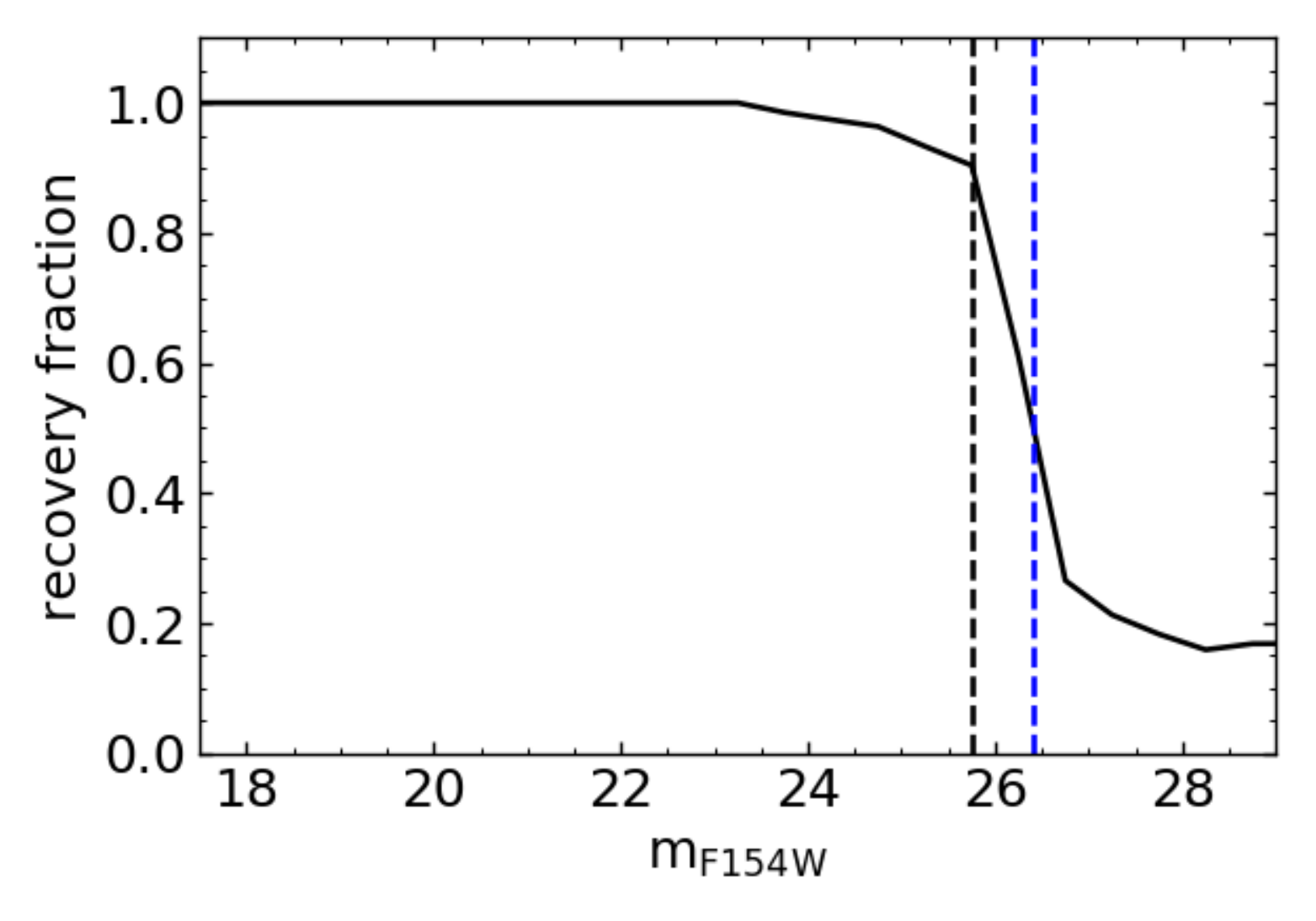}
    \includegraphics[width=0.33\textwidth]{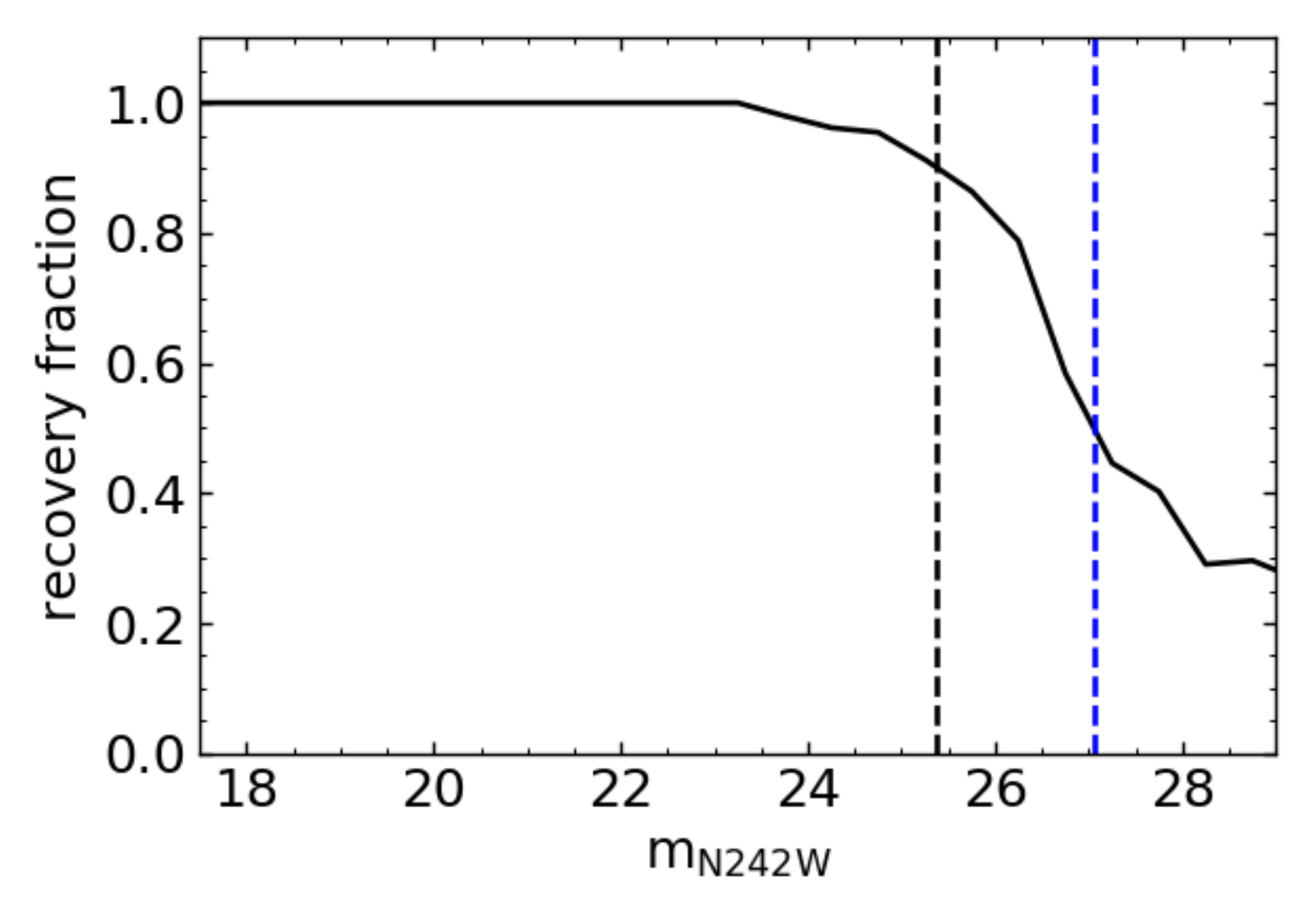} 
    \caption{The recovery fraction of injected simulated sources as a function of input magnitude for the FUV F154W (\textit{left}) and NUV N242W (\textit{right}) filter images. The 90\% and 50\% completeness limits for each image are marked in black and blue respectively. }
    \label{fig_recovery}
\end{figure*}

\begin{figure*}
    \centering
    \includegraphics[width=0.32\textwidth]{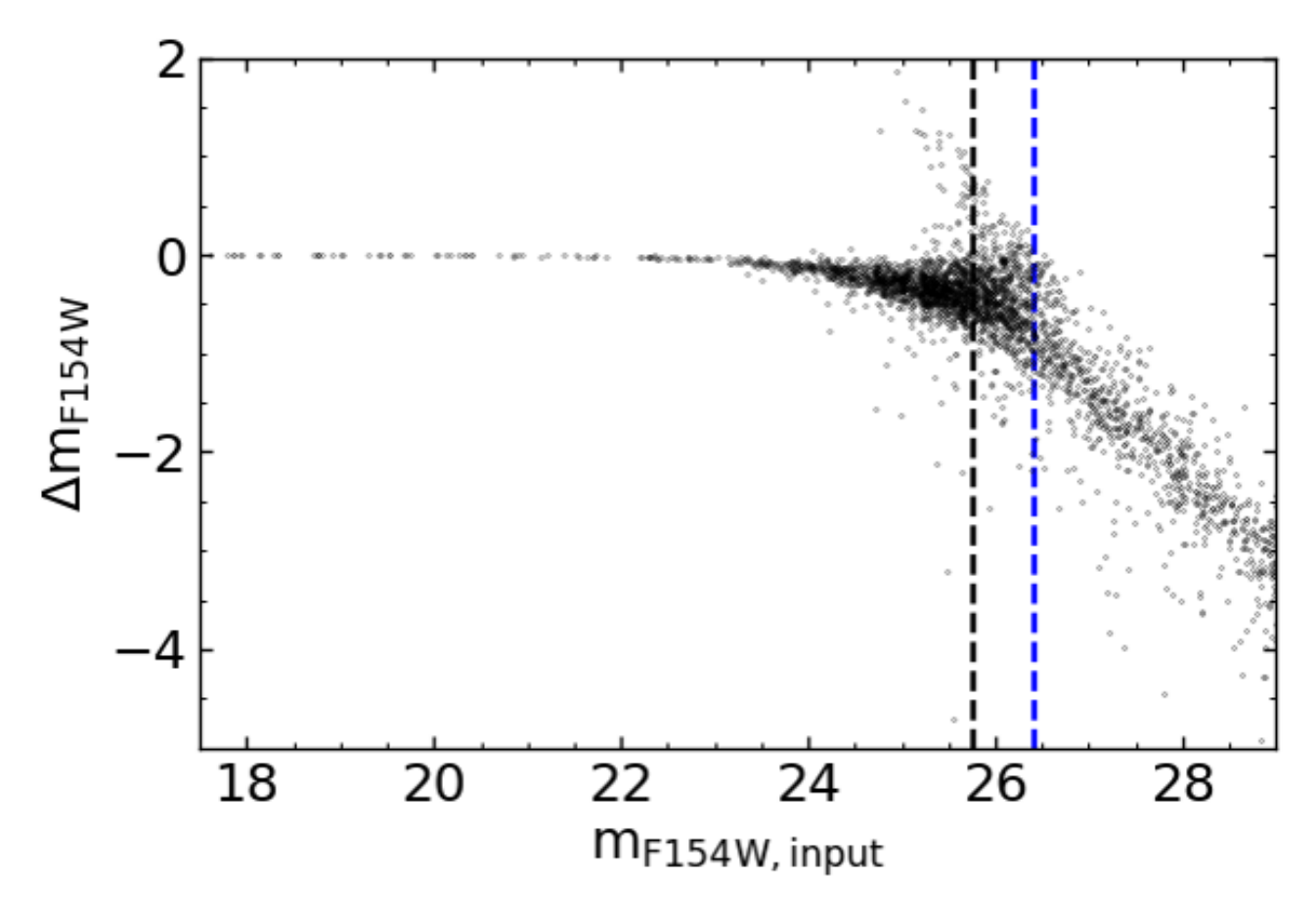}
    \includegraphics[width=0.32\textwidth]{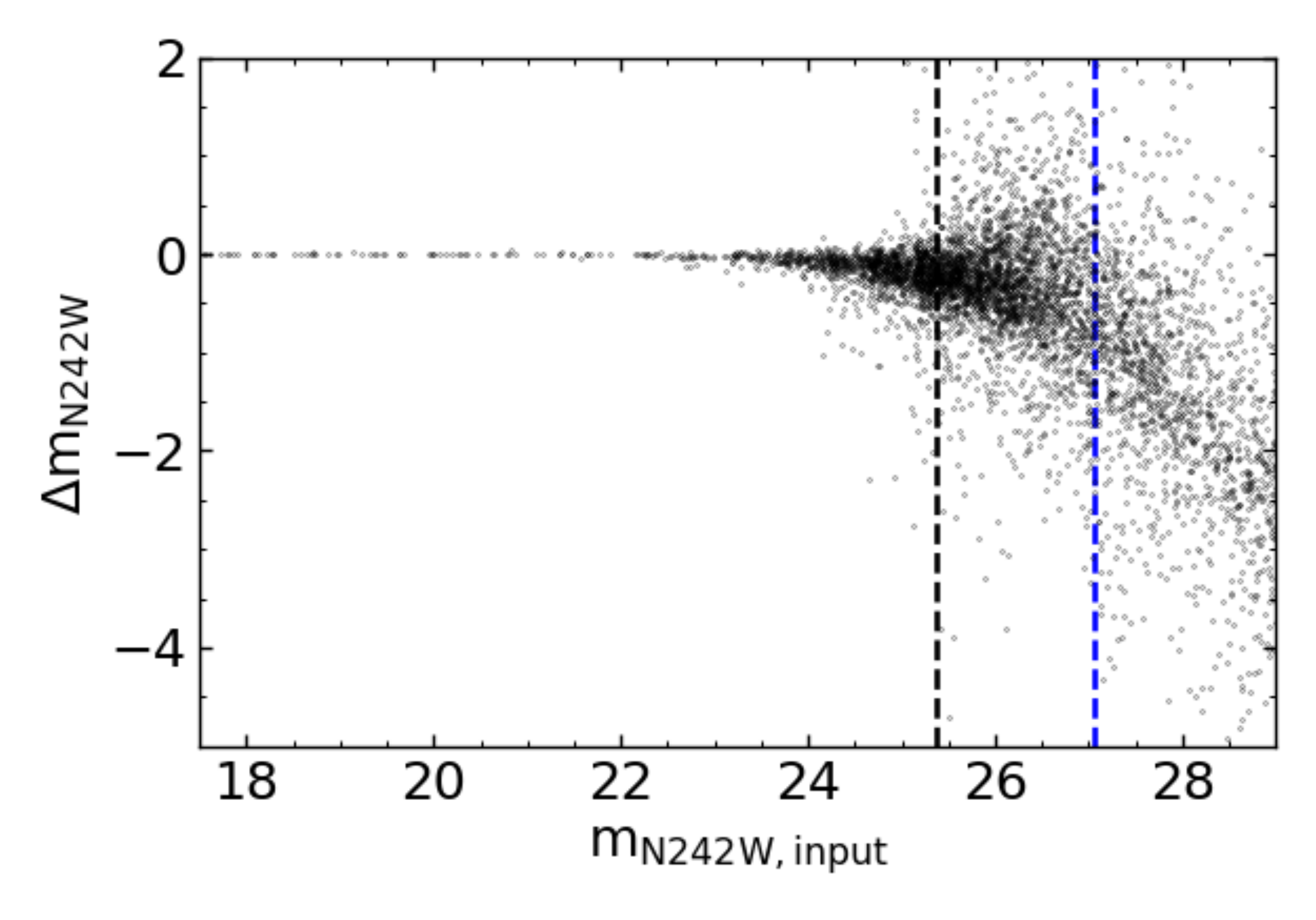} 
    \includegraphics[width=0.32\textwidth]{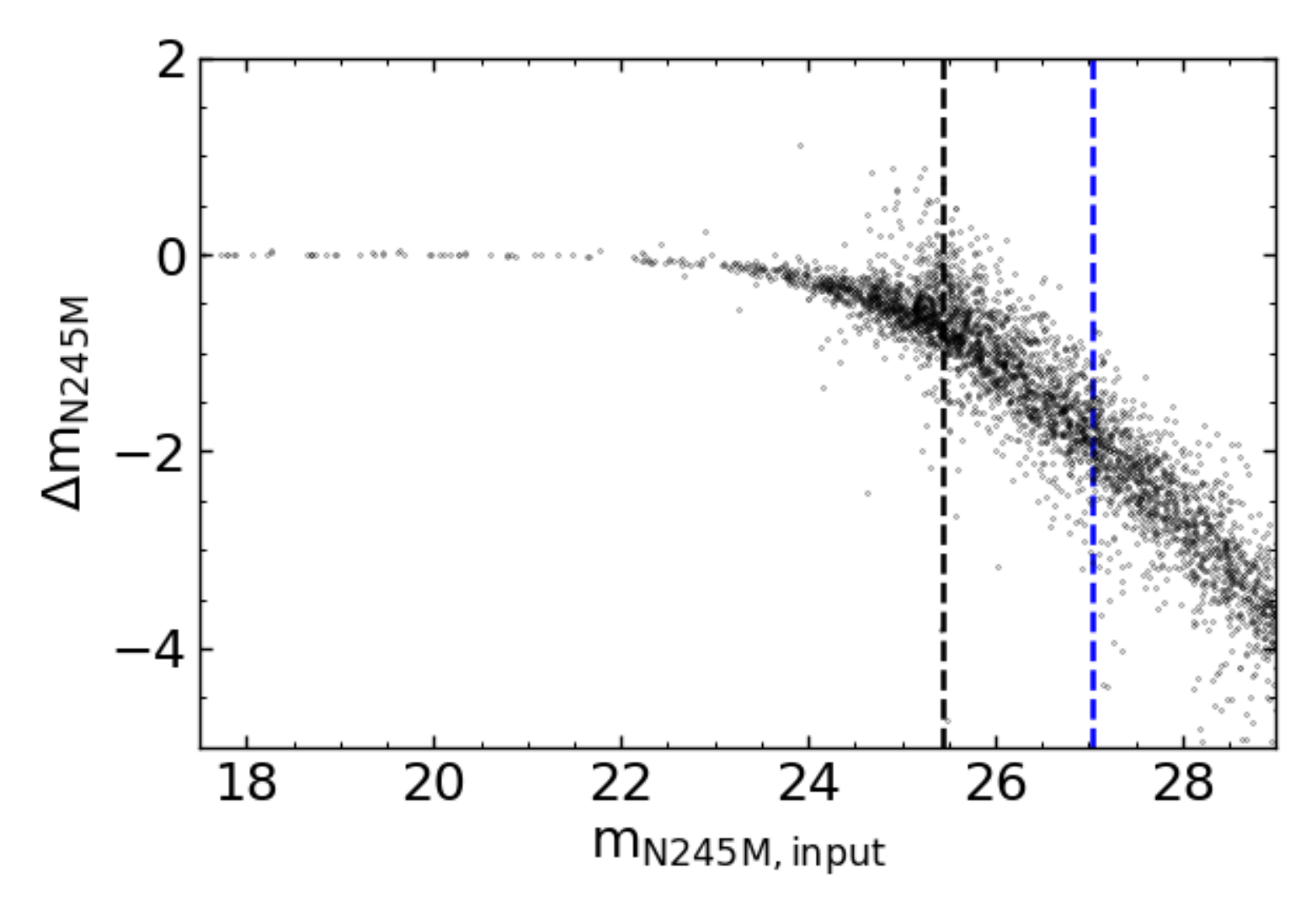} 
    \caption{The offset of measured magnitude from the input magnitude of injected simulated sources as a function of input magnitude for the FUV F154W (\textit{left}), NUV N242W (\textit{middle}) and NUV N245M (\textit{right}) filter images. The 90\% and 50\% completeness limits for each image are marked in black and blue respectively.}
    \label{fig_offset}
\end{figure*}

\subsection{Completeness}
\label{sect:completeness}
We carried out a completeness analysis of the UVIT images through the injection and subsequent recovery of simulated sources, similar to the analysis carried out for optical images in \citet{bhattacharya19}. Simulated sources of varying magnitude (each following the moffat PSF parameters described in Table~\ref{table_psf}; magnitude is then curve-of-growth corrected) were injected into the three filter images at randomised positions using the IRAF \textit{mkobjects} routine. Source detection and extraction was carried out as described earlier in Section~\ref{s_detection}. For each image, the recovery fraction is high for brighter sources and reduced for fainter sources. Figure~\ref{fig_recovery} shows the recovery fraction of injected simulated sources as a function of input magnitude for the FUV F154W and NUV N242W images. The 50\% completeness limits for each filter image are marked and noted in Table~\ref{table_uvit_obs}. Note the photometry for the NUV N245M filter was carried out in dual-mode with the detection on the NUV N242W image, thus it has the same completeness limits as the NUV N242W image. 

We further calculate the offset of magnitudes measured by SExtractor (individual detection on each image) using the input magnitude from the simulated sources in each image. This offset is plotted as a function of input magnitude for the three images in Figure~\ref{fig_offset}. The figure shows that the recovered magnitude is close to the input value for all three images for sources brighter than the 90\% completeness limit. The recovered magnitude is still close to the input value for sources brighter than the 50\% completeness limit, though some sources in this range appear brighter than their input magnitudes. 

There are a number of sources, fainter than the 50\% completeness limit in F154W and N242W (close to the 90\% completeness limit for N245M), that appear brighter than their input magnitude with the brightening increasing for fainter sources. We discuss the likely reason for this. The source detection falls sharply beyond the 50\% detection limit (see Figure~\ref{fig_recovery}). Those few sources that are still detected likely have some background pixels above the mean, close to the source center which are counted as source pixels, thereby increasing the source flux. Hence there would be a preference for only those sources brighter than their input values to be detected beyond the 50\% detection limit. As we go to even fainter sources, the relative background contribution to source flux increases. Thus fainter sources appear brighter than their input magnitudes with this offset increasing with faintness, as seen in Figure~\ref{fig_offset}.

\begin{table*}
\centering
\caption{Content of the UV catalog for HST-detected clean sources.$^{*}$}
\label{table_cat_candels}
\begin{tabular}{p{1cm}p{4cm}p{12.5cm}}
\hline
Column number & Column title & Description\\\hline
1 & ID & Unique HST source ID as given in HST CANDELS/GOODS-N catalog\\
2-3 & RA, DEC & RA (J2000) and DEC (J2000) of the object as given in HST CANDELS/GOODS-N catalog\\
4-7 & F154W\_m1, F154W\_m1err & F154W magnitude (without aperture correction) and error estimated with circular \\
 & F154W\_m2, F154W\_m2err & apertures of radii 0\farcs7 and 1\farcs4\,.\\
8-11 & N242W\_m1, N242W\_m1err & N242W magnitude (without aperture correction) and error estimated with circular \\
 & N242W\_m2, N242W\_m2err & apertures of radii 0\farcs7 and 1\farcs4\,.\\ 
12-15 & N245M\_m1, N245M\_m1err & N245M magnitude (without aperture correction) and error estimated with circular \\
 & N245M\_m2, N245M\_m2err & apertures of radii 0\farcs7 and 1\farcs4\,.\\ 

\hline
\end{tabular}
$^{*}$ Aperture correction values for the given UVIT magnitudes are listed in Table \ref{table_aper_corr} 
\end{table*}

\subsection{Catalog of HST CANDELS counterpart}
\label{s_candels}

UVIT has a comparatively larger PSF than the HST (Figure \ref{fig_uvit_galex}). Due to this un-matched PSF, multiple nearby objects in HST could appear as a single source in UVIT. Our FUV and NUV catalogs certainly contain such objects where UVIT could not resolve multiple sources separated by smaller than its PSF size. As one of the primary goals of our catalog is to identify potential LyC leaking galaxies, it is important to establish unique source correspondence between UV detection and other catalog that has better angular resolution. To explore the overall number of clean source identification within the HST-covered GOODS-north field, we made use of the HST CANDELS/GOODS-N photometric catalog \citep{barro2019}. Among the 35,451 sources listed in the CANDELS/GOODS-N catalog, we searched for objects that do not have any other detection within 1\farcs4 ($\sim$ UVIT PSF) radius. We found 27,209 such objects in the HST CANDELS catalog that do not have source confusion within the UVIT PSF size. We cross-matched the UVIT detections with this clean HST source list using a 1\farcs4 match-radius. The resulting catalog, with sources brighter than the 50\% completeness limit, provides 1,245 and 4,148 matched-detection in FUV and NUV, respectively. Relative to the total UVIT detection within HST coverage, these numbers convey that a significant percentage (50\% in FUV and 68\% in NUV) of sources have been identified in UVIT where we find only one HST object within 1\farcs4 radius.

\begin{figure}
    \centering
    \includegraphics[width=3.5in]{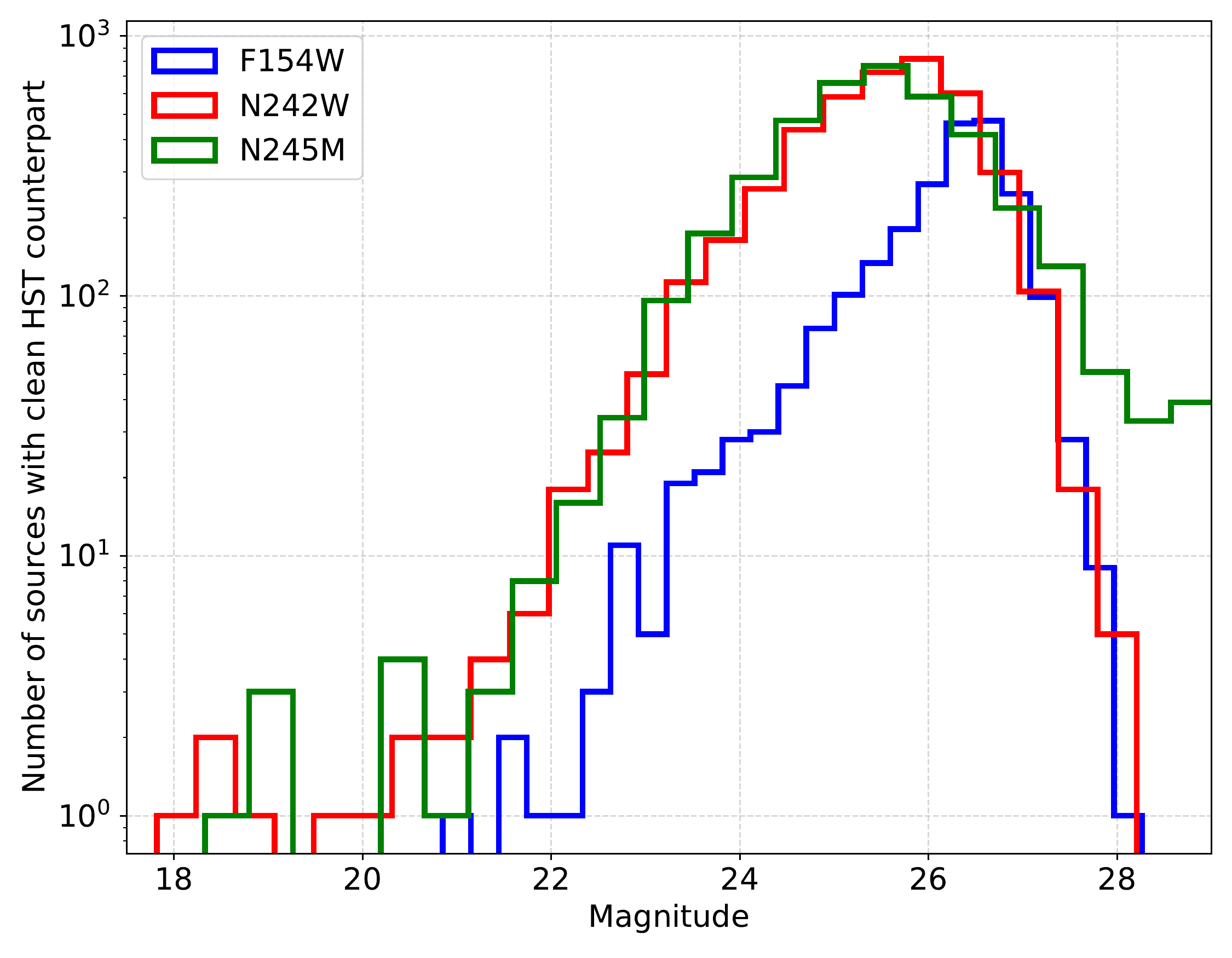} 
    \caption{The apparent magnitude (estimated with 1\farcs4 circular aperture with aperture correction) of UVIT identified objects that do not have source confusion within a 1\farcs4 radius as per the HST CANDELS/GOODS-N source catalog. These histograms are plotted for objects detected within the HST covered region of the UVIT field. F154W, N242W, and N245M filters are represented in blue, red, and green, respectively.}
    \label{fig_mag_single_hist}
\end{figure}

In Figure \ref{fig_mag_single_hist}, we show the magnitude of these cross-matched single objects for all three bands. The CANDELS catalog uses F160W near-infrared band to detect objects. As the UVIT filters sample the bluer part of the spectra, many of the HST objects can be undetected in UVIT images due to their negligible or lack of UV emission. In the case of high-redshift galaxies, UVIT filters could appear on the bluer side of Lyman break, resulting in a less chance to detect UV photons. Both of these factors can elude the source confusion in UVIT even if there exist multiple HST objects within the UVIT PSF. Hence, the number of clean detection noted here should be considered as the lower limit, which has a much higher value in reality.

Considering the availability of HST multi-band photometry of sources in the GOODS-north deep field, we produced an additional UV catalog listing the FUV and NUV magnitudes of the HST-detected sources. We used the positional prior of the 27,209 clean HST sources from the HST CANDELS/GOODS-N catalog \citep{barro2019} and performed fixed aperture photometry on the background-subtracted UVIT images using \textit{photutils} \citep{bradley2020} package. In the CANDELS catalog, including the measurement of total flux value for each detected object, \citet{barro2019} have also provided source-flux using multiple circular apertures. We considered two circular apertures of radii 0\farcs7 and 1\farcs4 (for which HST photometry is also available) and estimated magnitude in all the three UVIT bands. We provide photometry in two different apertures so that users can choose the optimal one depending on the immediate surrounding of a source. The UV magnitudes provided in the catalog are not aperture-corrected but we provide the correction values in Table \ref{table_aper_corr} for both the apertures. Utilizing the HST PSF images supplied by \citet{barro2019}, we also estimated aperture-correction values in the HST filters (listed in Table \ref{table_aper_corr}) for which photometry is available in the CANDELS catalog. Users can use both UVIT (from our catalog) and HST (from CANDELS catalog) photometry including the aperture-correction values, listed in Table \ref{table_aper_corr}, in either of the two apertures for color measurement or SED fitting of detected sources. One can also use the total flux of a source given in the CANDELS catalog along with the aperture-corrected UV flux for the same purpose. In our catalog, we provide the unique CANDELS source ID, source position (i.e., RA, DEC) from the HST catalog, and the UV photometry (without aperture-correction) in three UVIT filters. We have listed the content of this UV catalog produced for the HST-detected clean sources in Table \ref{table_cat_candels}. The catalog is made available in electronic format.

\section{Summary}

\label{s_summary}
We present deep UV images of the AUDFn (i.e., GOODS-N) field observed using three AstroSat/UVIT filters (F154W, N242W, and N245M). Our study provides the FUV and NUV source catalog of the AUDFn field which includes the entire HST covered GOODS-N region, notably enriching the multi-wavelength database of the field. The UV catalog fills an important gap in the already existing data from different telescopes. The inclusion of UVIT photometry on the bluer end of SEDs will help to better constrain the properties of many distant galaxies. This unique data set will further expand the search of LyC leaking galaxies, particularly in the intermediate redshift range \citep{saha2020}. Due to the scarcity of FUV imaging in deep fields, our analysis of the sky background and noise sheds new light on the understanding of the UV sky, which has a lower number photon events. The key outcome of this study are summarised below:

\begin{itemize}

    \item This work provides one of the first studies presenting deep field imaging of the AUDFn field using AstroSat/UVIT multi-band observation.
    
    \item With an exposure time of $\sim$\,34ks in F154W, 19ks in N242W, and 15ks in N245M, we reached a 3$\sigma$ detection limits of $\sim$ 27.35 mag, 27.28 mag and 27.02 mag, respectively. This highlights the well suited sensitivity of UVIT for conducting deep field surveys.
    
    \item We performed a careful inspection of the background mean and rms across the UVIT field to understand the nature of UV sky in deep field, especially in FUV. We find a $\sim$\,9\% variation in the mean background both in FUV and NUV across the field.
    
    \item We constructed PSFs for each band and find the FWHM as 1\farcs18, 1\farcs11, and 1\farcs24 in F154W, N242W, and N245M bands, respectively. The FWHM of stars selected to measure the N242W PSF ranges between 1\farcs0\,--\,1\farcs4 with a slight trend of increasing size with increasing distance from the detector centre. The ellipticity of these candidate stars ranges between 0.87\,--\,0.99.
    
    \item Using artificially injected sources, we found the 50\% completeness limits to be 26.40\,mag and 27.05\,mag in FUV and NUV, respectively. The sources included in the catalogs with magnitude brighter than this limit have more reliable photometry.
    
    \item We identify a total of 16001 (FUV) and 16761 (NUV) sources with 6839 and 16171 objects above the 50\% completeness limit in the respective bands. The number of sources detected only within the HST covered region is 6082 (F154W) and 6292 (N242W). Our detection finds that 48\% (FUV) and 64\% (NUV) of the sources brighter than the respective 50\% completeness limit had only one HST counterpart within a 1\farcs4 radius.
    
    \item We also provide UV photometry for 27,209 HST-detected clean sources using their positional prior from the CANDELS/GOODS-north catalog.
    
\end{itemize}

\acknowledgements

This work is primarily based on observations taken by AstroSat/UVIT. The UVIT project is a result of collaboration between IIA, Bengaluru, IUCAA, Pune, TIFR, Mumbai, several centres of ISRO, and CSA. Indian Institutions and the Canadian Space Agency have contributed to the work presented in this paper. Several groups from ISAC (ISRO), Bengaluru, and IISU (ISRO), and Trivandrum have contributed to the design, fabrication, and testing of the payload. The Mission Group (ISAC) and ISTRAC (ISAC) continue to provide support by making observations with, and reception and initial processing of the data. RAW and RAJ acknowledge support from NASA JWST Interdisciplinary Scientist grants NAG5-12460, NNX14AN10G and 80NSSC18K0200 from GSFC. We gratefully thank all the individuals involved in the various teams for providing their support to the project from the early stages of the design to launch and observations with it in the orbit. We acknowledge support from HST grants HST-GO-15647, and we used observations taken by the CANDELS survey (HST-GO-9425 and HST-GO-9583) with the NASA/ESA Hubble Space Telescope, which is operated by the Association of Universities for Research in Astronomy, Inc., under NASA contract NAS5-26555. This research made use of Matplotlib \citep{matplotlib2007}, Astropy \citep{astropy2013,astropy2018}, photutils \citep{bradley2020} community-developed core Python packages for Astronomy and SAOImageDS9 \citep{joye2003}. Finally, we thank the referee for valuable suggestions.

\software{SExtractor \citep{bertin1996}, CCDLAB \citep{postma2017}, SAOImageDS9 \citep{joye2003}, Matplotlib \citep{matplotlib2007}, Astropy \citep{astropy2013,astropy2018}, photutils \citep{bradley2020}}


\end{document}